\documentclass[reprint,
 amssymb, amsmath,%
 aip,cha,%
groupedaddress,%
frontmatterverbose,
]{revtex4-1}  

\usepackage{graphicx}
\usepackage{amsmath,amssymb,amsfonts,mathtools}
\usepackage{ulem}
\usepackage{color}
\usepackage{multirow}
\usepackage{float}
\usepackage{csquotes}

\begin{document}

\title{State Dependent Ring Polymer Molecular Dynamics for Investigating Excited Nonadiabatic Dynamics }
\author{Sutirtha N. Chowdhury}
\author{Pengfei Huo}
\email{pengfei.huo@rochester.edu}
\affiliation{Department of Chemistry, University of Rochester, 120 Trustee Road, Rochester, New York 14627, United States}%

\begin{abstract}
Recently proposed non-adiabatic ring polymer molecular dynamics (NRPMD) approach has shown to provide accurate quantum dynamics by incorporating explicit electronic state descriptions and nuclear quantizations. Here, we present a rigorous derivation of the NRPMD Hamiltonian and investigate its performance on simulating excited state non-adiabatic dynamics. Our derivation is based on the Meyer-Miller-Stock-Thoss (MMST) mapping representation for electronic states and the ring-polymer path-integral description for nuclei, resulting in the same Hamiltonian proposed in the original NRPMD approach. In addition, we investigate the accuracy of using NRPMD to simulate photoinduced non-adiabatic dynamics in simple model systems. These model calculations suggest that NRPMD can alleviate the zero-point energy leakage problem that is commonly encountered in the classical Wigner dynamics, and provide accurate excited states non-adiabatic dynamics. This work provides a solid theoretical foundation of the promising NRPMD Hamiltonian and demonstrates the possibility of using state-dependent RPMD approach to accurately simulate electronic non-adiabatic dynamics while explicitly quantize nuclei. 
\end{abstract}

\maketitle
\section{Introduction}
Accurately and efficiently simulating electronic non-adiabatic transitions and nuclear quantum effects remain one of the central challenges in theoretical chemistry.\cite{tully2012,althorpe2016} Directly performing exact quantum dynamics simulations in large systems are computationally demanding, despite exciting recent progress.\cite{Meyer1990,beck2000,thoss2003,Burghardt2018,Habershon2018,Shalashilin,Martinez, ishizaki05,tanimura06,yan09,Schulten, makri95,makri95jcp,keeling2017,Walters2016JCP,Walters:2015,batista2017,chan18} It is thus ideal to develop trajectory-based approximate methods that can accurately describe the electronic and nuclear quantum effects. To this end, a large number of these approaches are developed, including the popular trajectory surface-hopping method (mixed quantum-classical approach),\cite{tully1990,subotnik2016,wang2016,Barbatti} the semi-classical path-integral approaches,\cite{miller2001,miller2009,sun1998,shi2004,bonella2005, makri2011, huo2011,lee2016} the mixed quantum-classical Liouville equation, \cite{kernan2008,kim2008,Hsieh2012,Hsieh2013,Kapral2015} the symmetrical quasi-classical (SQC) approach,\cite{Miller:2016b,tao2016,Geva2018,Sandoval2018,cotton2019} and the generalized quantum master equation (GQME).\cite{Geva2006,Markland2015,Markland2015jcp,Markland2016,Geva2019} Despite that they are promising to accurately describe electronic non-adiabatic dynamics, these approaches generally do not preserve quantum Boltzmann distribution (QBD)\cite{parandekar2005,schmidt2008} or zero point energy (ZPE) associated with the nuclear degrees of freedom (DOF), and often suffer from numerical issues such as ZPE leakage, \cite{habershon2009,muller1999} although significant improvements are accomplished through the SQC\cite{Miller:2016b} and GQME approaches.\cite{Markland2015jcp,Markland2016,Geva2019}

Imaginary-time path-integral approaches,\cite{berne1986,ceperley1995,chandler1981} such as the centroid molecular dynamics (CMD)\cite{cao1994,jang1999} and the ring-polymer molecular dynamics (RPMD),\cite{habershon2013,craig2004} resemble classical MD in an extended phase space and provide a convenient way to compute approximate quantum time-correlation functions.\cite{habershon2013} The classical evolution of RPMD preserves its initial quantum distribution captured by the ring-polymer Hamiltonian, and it is free of the zero-point energy leaking problem.\cite{habershon2009,habershon2013} Despite its success on describing quantum effects in the condensed phase, RPMD is limited to one-electron non-adiabatic dynamics\cite{menzeelev2010,menzeelev2011,kretchmer2013,kretchmer2016,Ananth2016} or nuclear quantization,\cite{habershon2013,boekelheide2011,markland2014,Markland:2017,Ceriotti:2018} as well as the lack of real-time electronic coherence effects.\cite{menzeelev2010,menzeelev2011} Recently emerged state-dependent RPMD approaches, such as non-adiabatic RPMD (NRPMD),\cite{richardson2013,hele2016, richardson2017} mapping variable RPMD (MV-RPMD),\cite{ananth2013,duke2015,ananth2017} Kinetically-constrained RPMD (KC-RPMD),\cite{menzeleev2014,kretchmer2016,kretchmer2018} coherent state RPMD (CS-RPMD),\cite{chowdhury2017} and ring-polymer surface hopping (RPSH) \cite{shushkov2012,shakib2017,miller2018,miller2019} are promising to provide accurate non-adiabatic dynamics with an explicit description of electronic states, in addition to the reliable treatment of nuclear quantum effects through ring polymer quantization. 

Among these state-dependent RPMD approaches, NRPMD\cite{richardson2013,richardson2017} and CS-RPMD\cite{chowdhury2017} have shown to accurately describe both the electronic dynamics and nuclear quantum effects. The NRPMD\cite{richardson2013,richardson2017} and CS-RPMD\cite{chowdhury2017} Hamiltonian can be viewed as the generalization of the Meyer-Miller-Stock-Thoss (MMST) mapping Hamiltonian\cite{meyer1979,stock1997,stock1999} with the ring-polymer description of the nuclei. Both Hamiltonians have many desired properties, such as a clear adiabatic limit (that returns to the original RPMD Hamiltonian) and one bead limit (that returns to the original MMST Hamiltonian), and recovering the correct electronic Rabi oscillations when the electronic states and nuclei are decoupled.\cite{richardson2013,richardson2017,chowdhury2017} Nevertheless, the promising NRPMD Hamiltonian is {\it proposed} through a physically motivated but {\it ad hoc} fashion.\cite{richardson2013}

In this paper, we provide a rigorous derivation of the NRPMD Hamiltonian, which is based on the MMST mapping formalism\cite{meyer1979,stock1997,stock1999} for the electronic DOF and the ring polymer path-integral representation\cite{feynman1965,berne1986,ceperley1995,chandler1981} for the nuclear DOF, leading to the same Hamiltonian that has been previously proposed in the NRPMD approach.\cite{richardson2013,richardson2017} The NRPMD Hamiltonian and the previously derived CS-RPMD Hamiltonian\cite{chowdhury2017} can be viewed as a unified classical theory for electronic states (through the MMST mapping formalism) and nuclei (through the ring polymer quantization). 

In addition, we perform numerical simulations to investigate the accuracy of NRPMD Hamiltonian for excited states non-adiabatic dynamics. Despite that RPMD-based approaches are initially developed for investigating quantum dynamics under thermal-equilibrium conditions,\cite{habershon2013} recent work\cite{welsch2016} based on the Matsubara dynamics framework\cite{hele2015,hele2015jcp2} (which does not subject to the restriction of equilibrium conditions\cite{welsch2016}) has demonstrated that RPMD yields reliable non-equilibrium time-correlation function. Further, RPMD based approaches\cite{menzeelev2010,duke2015} have already been used to simulate non-adiabatic dynamics under non-equilibrium initial conditions. These studies inspire us to investigate the performance of NRPMD for simulating excited states non-adiabatic dynamics. 

Our numerical results with model calculations suggest that NRPMD can provide an accurate short-time non-adiabatic branching dynamics among many coupled electronic states, as well as the longer time dynamics such as the plateau value or recurrence of the oscillations of the electronic population. Quantizing nuclei through ring polymer instead of using the Wigner distribution alleviates zero-point energy leakage problems that classical Wigner dynamics encountered, leading to more accurate electronic quantum dynamics. These encouraging numerical results, together with our rigorous derivation of the NRPMD Hamiltonian open up new possibilities to accurately simulate non-adiabatic quantum dynamics while explicitly quantize nuclei. 

\section {Theory}
In this section, we provide a rigorous derivation of the NRPMD Hamiltonian\cite{richardson2013,richardson2017} through the imaginary-time formalism. Based upon that, we propose a numerical procedure to compute the electronic population for excited state non-adiabatic dynamics. We start by expressing the total Hamiltonian operator as 
\begin{equation}\label{eqn:tot-H}
\hat{H}=\hat{T}+\hat{V}_{0}+\hat{H}_{e} ={\hat{\bf P}^{2}\over{2{\bf M}}}+V_{0}({\bf \hat{R}})+\sum_{n,m=1}^\mathcal{K}V_{nm}({\bf \hat{R}})|n\rangle \langle m|,
\end{equation}
where $\{|n\rangle\}$ is the {\it diabatic} basis, $\hat{T}$ is the nuclear kinetic energy operator, $\hat{\bf R}  \equiv \{\hat{R}_{1},...,\hat{R}_\mathcal{F}\}$ is the nuclear position operator associated with $\mathcal{F}$ different nuclear DOF, with the corresponding conjugate momentum operator $\hat{\bf P}\equiv \{\hat{P}_{1},...,\hat{P}_\mathcal{F}$\} and the nuclear mass ${\bf M}\equiv\{M_1, M_2, ...,M_\mathcal{F}\}$. In addition, $V_{0}({\bf \hat{R}})$ is the state-independent potential operator, whereas $\hat{H}_\mathrm{e}=\sum_{nm}V_{nm}({\bf \hat{R}})|n\rangle \langle m|$ is the state-dependent potential operator (electronic part of the Hamiltonian) with $\mathcal{K}$ total diabatic electronic states. 

The canonical partition function of the system is defined as $\mathcal{Z}=\mathrm{Tr_{en}}[e^{-\beta\hat H}]$, where $\mathrm{Tr_{en}=Tr_{e}Tr_{n}}$ represents the trace over both the electronic and nuclear DOFs, with $\mathrm{Tr_{e}}[...]\equiv \sum_{n=1}^{\mathcal{K}}\langle n|...|n\rangle$, and $\mathrm{Tr_{n}}[...]\equiv \int (...)d {\bf R}$. Further, $\beta=1/k_\mathrm{B}T$ is the reciprocal temperature, and $\hat{H}$ is the total Hamiltonian operator defined in Eqn.~\ref{eqn:tot-H}. The partition function can be exactly evaluated as $\mathcal{Z}=\mathrm{Tr_{en}}\left[\prod_{\alpha = 1}^{N}e^{-\beta_{N}\hat H}\right]$, with a higher effective temperature defined as $\beta_{N}=\beta/N$. Further splitting the Boltzmann operator by trotter expansion under the infinite bead limit $N\rightarrow\infty$ gives $\mathcal{Z}=\lim_{N\to\infty}\mathrm{Tr_{en}}\left[ \prod_{\alpha = 1}^{N}e^{-\beta_{N}(\hat{T}+\hat{V}_{0})}e^{-\beta_{N}\hat{H}_{e}}\right]$. Inserting $N$ copies of the resolution of identity $I_{\bf R} = \int {\bf dR_{\alpha}} |\bf {R_{\alpha}}\rangle\langle {\bf R_{\alpha}}|$ and $I_{\bf P}= \int {\bf dP_{\alpha}} |\bf {P_{\alpha}}\rangle\langle {\bf P_{\alpha}}|$, and explicitly performing the trace over the nuclear DOF based on the standard path-integral technique, \cite{feynman1965,berne1986,ceperley1995,chandler1981} we have 
\begin{equation}\label{eqn:part-gen}
\mathcal{Z}  = \lim_{N\to\infty} \int d \{{\bf P_\alpha}\} d\{{\bf R_\alpha}\} e^{-\beta_{N}H^{0}_N}\text{Tr}_\mathrm{e}\left[\prod_{\alpha = 1}^{N} e^{-\beta_{N}\hat {H}_{e}({\bf R_{\alpha}})}\right],
\end{equation}                                                      with $\int d \{{\bf P_\alpha}\} d\{{\bf R_\alpha}\}=\int \prod_{\alpha = 1}^{N}  d {\bf P_\alpha} d{\bf R_\alpha}$. The trace over electronic DOF can be performed with diabatic states as $\mathrm{Tr}_\mathrm{e}[...]=\sum_{n=1}^\mathcal{K}\langle n|...|n\rangle$. The state-independent Hamiltonian $H^{0}_{N}$ is expressed as follows
\begin{equation}\label{eqn:hrp}
H^{0}_{N}=\sum_{\alpha=1}^{N}{{{\bf P_{\alpha}}^{2}}\over{2{\bf M}}}+{{\bf M}\over{2\beta^{2}_{N}\hbar^{2}}}({\bf R_{\alpha}-R_{\alpha-1}})^{2}+V_{0}({\bf R_{\alpha}}), 
\end{equation}
where $H_\mathrm{rp}=H^{0}_{N}-\sum_{\alpha}V_0({\bf R_{\alpha}})$ is so-called the free ring-polymer Hamiltonian.\cite{craig2004,ceriotti2010}
The above partition function is a common expression for all state-dependent RPMD approaches\cite{menzeleev2014,ananth2013,richardson2013,hele2011,duke2016} and PIMC methods.\cite{alexander2001,schmidt2007,ananth2010, lu2017} The only difference among these approaches arises from the treatment of the electronic term $\text{Tr}_\mathrm{e}[\prod_{\alpha = 1}^{N}e^{-\beta_{N}\hat {H}_{e}({\bf R_{\alpha}})}]$. For example, in the mean-field RPMD approach,\cite{hele2011,duke2016} the electronic potential is obtained from a weighted average of ring-polymer in different electronic configurations; in the MV-RPMD approach,\cite{ananth2013} the electronic states are explicitly described with mapping variables in the Wigner representation.

In this work, we adapt the MMST mapping representation\cite{meyer1979,stock1997,stock1999} to transform discrete electronic states into continuous variables. Based on this representation, $\mathcal{K}$ diabatic electronic states are mapped onto $\mathcal{K}$ harmonic oscillators' ground and first excited states through the following relation 
\begin{equation}
| n \rangle \rightarrow |0_{1}...1_{n}...0_\mathcal{K} \rangle=\hat{a}^{\dagger}_{n}|0_{1}...0_{n}...0_\mathcal{K} \rangle.
\end{equation}
Here, $|0_{1}...1_{n}...0_\mathcal{K} \rangle$ is the singly excited oscillator (SEO) state with $(\mathcal{K}-1)$ oscillators in their ground states and the $n_\mathrm{th}$ oscillator in its first excited state. Thus, the MMST formalism provides\cite{stock1997,stock1999} the following mapping relation 
\begin{equation}
| n \rangle \langle m| \rightarrow\hat{a}^{\dagger}_{n}\hat{a}_{m},
\end{equation}
with $\hat{a}^{\dagger}_{n}={1/{\sqrt{2\hbar}}}\left(\hat{q}_{n}-\mathrm{i}\hat{p}_{n}\right)$ and $\hat{a}_{m}={1/{\sqrt{2\hbar}}}\left(\hat{q}_{m}+\mathrm{i}\hat{p}_{m}\right)$ as the creation and annihilation operators of the harmonic oscillator. 

With the MMST representation, we express the state-dependent potential operator $\hat {H}_\mathrm{e}({\bf R_{\alpha}})$ in Eqn.~\ref{eqn:part-gen} as follows \begin{equation}\label{eqn:state_dep_pot}
\hat {H}_\mathrm{e}({\bf R_{\alpha}}) \equiv \sum_{n,m}V_{nm}({\bf R_\alpha})|n\rangle\langle m|  \rightarrow \sum_{n,m}V_{nm}({\bf R_\alpha})\hat a_{n}^{\dagger}\hat a_{m}.
\end{equation}
This is commonly referred as the MMST Hamiltonian.\cite{meyer1979,stock1997}  Using the above MMST representation, we can derive a partition function that contains the NRPMD Hamiltonian.

\subsection{Derivation of the NRPMD Hamiltonian} 
We begin by replacing the trace over the electronic DOF in Eqn.~\ref{eqn:part-gen} with the phase space integral of mapping variables in the Wigner representation\cite{case08,hele2016,ananth2013,duke2015} as follows
\begin{eqnarray}\label{eqn:elec_part}
&&\text{Tr}_\mathrm{e}\big[\prod_{\alpha = 1}^{N} e^{-\beta_{N} \hat{H}_{e}({\bf R}_\alpha)}\big] \\
&&=\frac{1}{(2\pi\hbar)^\mathcal{K}}\int d{\bf q}_{1}d{\bf p}_{1}\big[e^{-\beta_{N} \hat{H}_{e}({\bf R}_{1})} \prod_{\alpha = 2}^{N} e^{-\beta_{N} \hat{H}_{e}({\bf R}_\alpha)}\mathcal {P}\big]_{\mathrm{w}_{1}}.\nonumber
\end{eqnarray}
In the above equation, the Wigner-Weyl transform\cite{Wigner1932,Hillery1984,case08} of the $\alpha_\mathrm{th}$ bead's mapping DOF is defined as 
\begin{equation}\label{eqn:wigner}
[{O}]_{\mathrm{w}_\alpha} =  \int d{\bf \Delta}_{\alpha}e^{i{\bf p}_{\alpha}^\mathrm{T}{\bf \Delta}_{\alpha}/\hbar}\left\langle {\bf q}_{\alpha}-\frac{{\bf \Delta}_\alpha}{2}\left|\mathrm{\hat O}\right|{\bf q}_{\alpha}+\frac{{\bf \Delta}_{\alpha}}{2}\right\rangle.
\end{equation} 
We use the notation ${\bf q}_{\alpha}\equiv\{[{\bf q}_{\alpha}]_{1},...[{\bf q}_{\alpha}]_{n},...[{\bf q}_{\alpha}]_{\mathcal{K}}\}$ and ${\bf p}_{\alpha} \equiv \{[{\bf p}_{\alpha}]_{1},...[{\bf p}_{\alpha}]_{n},...[{\bf p}_{\alpha}]_{\mathcal{K}}\}$ to represent $\mathcal{K}$ mapping variables for $\mathcal{K}$ electronic states associated with the $\alpha_\mathrm{th}$ imaginary-time slice. Further, we have inserted an electronic projection operator $\mathcal {P}=\sum_{n}|n\rangle\langle n|$ to restrain the mapping variables within the SEO subspace.\cite{ananth2013,ananth2010}

Recall that the Wigner transform has the following property\cite{case08} 
\begin{equation}\label{eqn:wig-prop}
\int d{\bf q}d{\bf p} [AB]_\mathrm{w}=\int d{\bf q}d{\bf p} [A]_\mathrm{w}[B]_\mathrm{w}, 
\end{equation}
where $[AB]_\mathrm{w}$,$[A]_\mathrm{w}$, and $[B]_\mathrm{w}$ are Wigner transforms of operator $\hat{A}\hat{B}$, $\hat{A}$, and $\hat{B}$, respectively, defined in Eqn.~\ref{eqn:wigner}. With the above equality, we can rewrite Eqn.~\ref{eqn:elec_part} as follows
\begin{eqnarray}\label{eqn:power_series}
&&\text{Tr}_\mathrm{e}\bigg[\prod_{\alpha = 1}^{N} e^{-\beta_{N} \hat{H}_\mathrm{e}({\bf R}_\alpha)}\bigg]  \\
&&=\frac{1}{(2\pi\hbar)^\mathcal{K}}\int d{\bf q}_{1}d{\bf p}_{1}\bigg[e^{-\beta_{N} \hat{H}_\mathrm{e}({\bf R}_{1})} \prod_{\alpha = 2}^{N} e^{-\beta_{N} \hat{H}_\mathrm{e}({\bf R}_\alpha)}\mathcal {P}\bigg]_{\mathrm{w}_1} \nonumber  \\
&&=\frac{1}{(2\pi\hbar)^\mathcal{K}}\int d{\bf q}_{1}d{\bf p}_{1}\bigg[e^{-\beta_{N}\hat{H}_\mathrm{e}({\bf R}_{1})}\bigg]_{\mathrm{w}_1}\bigg[\prod_{\alpha = 2}^{N} e^{-\beta_{N}\hat{H}_{e}({\bf R}_\alpha)}\mathcal {P}\bigg]_{\mathrm{w}_1}\nonumber \\
&&=\frac{1}{(2\pi\hbar)^\mathcal{K}}\int d{\bf q}_{1}d{\bf p}_{1} \bigg[1-\beta_{N} \hat{H}_\mathrm{e}({\bf R}_{1})\bigg]_{\mathrm{w}_1} \bigg[\prod_{\alpha = 2}^{N} e^{-\beta_{N}\hat{H}_{e}({\bf R}_\alpha)}\mathcal {P}\bigg]_{\mathrm{w}_1}\nonumber \\
&&=\frac{1}{(2\pi\hbar)^\mathcal{K}}\int d{\bf q}_{1}d{\bf p}_{1} e^{-\beta_N [\hat H_{e}({\bf R}_{1})]_{\mathrm{w}_1}} \left[\prod_{\alpha = 2}^{N} e^{-\beta_{N}\hat{H}_\mathrm{e}({\bf R}_\alpha)}\mathcal {P}\right]_{\mathrm{w}_1}  \nonumber
\end{eqnarray}
To derive the last two lines of the above equation, we use the fact that under the limit $\beta_N\rightarrow 0$, Boltzmann operator becomes $e^{-\beta_{N} \hat{H}_{e}({\bf R}_{1})}=1-\beta_{N} \hat{H}_{e}({\bf R}_{1})+\mathcal{O}(\beta_{N}^{2})$. Under this limit, $[e^{-\beta_{N} \hat{H}_{e}({\bf R}_{1})}]_{\mathrm{w}_1}=[1-\beta_{N} \hat{H}_{e}({\bf R}_{1})]_{\mathrm{w}_1}=1-\beta_{N}[\hat{H}_{e}({\bf R}_{1})]_{\mathrm{w}_1}$. Expressing this term back to the full exponential factor, we arrived at the last line of Eqn.~\ref{eqn:power_series}.

To evaluate $[\hat{H}_\mathrm{e}({\bf R}_{1})]_{\mathrm{w}_1}$, we use the mapping Hamiltonian relation described in Eqn.~\ref{eqn:state_dep_pot} and the Wigner transform defined in Eqn.~\ref{eqn:wigner}, leading to the following expression
\begin{eqnarray}\label{eqn:he_r1}
&&[{\hat H}_\mathrm{e}({\bf R}_{1})]_\mathrm{w_1}\\
&=&\int d{\bf \Delta}_{1} e^{i{\bf p}_{1}^\mathrm{T}{\bf \Delta}_{1}/\hbar}\langle {\bf q}_1-\frac{{\bf \Delta}_1}{2} |\sum_{nm}V_{nm}({\bf R}_{1})\hat {a}_{n}^{\dagger}\hat{a}_{m} | {\bf q}_1+\frac{{\bf \Delta}_1}{2} \rangle \nonumber
\end{eqnarray}
To obtain an explicit expression, we use $\hat{a}^{\dagger}_{n}={1/{\sqrt{2\hbar}}}\left(\hat{q}_{n}-\mathrm{i}\hat{p}_{n}\right)$ and $\hat{a}_{m}={1/{\sqrt{2\hbar}}}\left(\hat{q}_{m}+\mathrm{i}\hat{p}_{m}\right)$, and evaluate these Wigner integrals. This derivation is provided in Appendix A, giving the Wigner transformed MMST mapping Hamiltonian of the $\alpha_\mathrm{th}$ bead as follows
\begin{eqnarray}\label{eqn:mapwig}
&&[{\hat H}_{e}({\bf R}_\alpha)]_{\mathrm{w}_\alpha}\\
&&= \frac{1}{2\hbar}\sum_{nm}V_{nm}({\bf R}_\alpha)\bigg([{\bf q}_{\alpha}]_{n}[{\bf q}_{\alpha}]_{m}+ [{\bf p}_{\alpha}]_{n}[{\bf p}_{\alpha}]_{m}-\delta_{nm}\hbar\bigg)\nonumber.
\end{eqnarray}

To evaluate $\left[\prod_{\alpha = 2}^{N} e^{-\beta_{N}\hat{H}_{e}({\bf R}_\alpha)}\mathcal {P}\right]_{\mathrm{w}_1}$ in Eqn~\ref{eqn:power_series}, we repeat the same procedure used in Eqn.~\ref{eqn:power_series} for the rest of the imaginary-time slices. The details of this derivation are provided in Appendix A, and here, we briefly summarize the procedure. First, we insert the resolution of identity $\int d {\bf q}_{2} \mathcal {P} |{\bf q}_2\rangle\langle {\bf q}_{2}|$ into $\left[\prod_{\alpha = 2}^{N} e^{-\beta_{N}\hat{H}_{e}({\bf R}_\alpha)}\mathcal {P}\right]_{\mathrm{w}_1}$. Second, we rearrange the order of terms in the integral and reexpress it as a trace over ${\bf q}_{2}$. Finally, replacing the trace over ${\bf q}_{2}$ by a Wigner transform, we can use the property in Eqn.~\ref{eqn:wig-prop} again and repeat the same procedure outlined in Eqn.~\ref{eqn:power_series} to factorize the total Wigner function as a product of two Wigner functions inside the $\int d{\bf q}_2 d{\bf p}_2$ integral. Repeating the above outlined process for every single bead, we ended up with the following expression of the electronic trace
\begin{eqnarray}\label{eqn:elec-part}
&&~~~\text{Tr}_\mathrm{e}\left[\prod_{\alpha = 1}^{N} e^{-\beta_{N} \hat{H}_\mathrm{e}({\bf R}_\alpha)}\right]\\
&&~~~=\frac{1}{(2\pi\hbar)^{N\mathcal{K}}}\int d\{{\bf q}_{\alpha}\} d \{{\bf p}_{\alpha}\}e^{-\beta_N \sum_{\alpha}[{\hat H}_{e}({\bf R}_\alpha)]_{\mathrm{w}_\alpha}} \nonumber\\
&&~~~~\times\int d\{{\bf \Delta}_{\alpha}\}\prod_{\alpha = 1}^{N}    e^{i{\bf p}_{\alpha}^\mathrm{T}{\bf \Delta}_{\alpha}/\hbar}  \langle {\bf q}_{\alpha}-\frac{{\bf \Delta}_\alpha}{2}| \mathcal {P} |{\bf q}_{\alpha+1}+\frac{{\bf \Delta}_{\alpha+1}}{2}\rangle. \nonumber
\end{eqnarray}

Analytically evaluate the last term of the above expression\cite{ananth2013} (with details provided in Appendix A), and plugging the result into Eqn.~\ref{eqn:part-gen}, we arrive at the final expression of the partition function
\begin{equation}\label{eqn:z_wigner}
\mathcal{Z}_{{N}}=\lim_{N\to\infty} \frac{1}{(2\pi\hbar)^{N\mathcal{K}}} \int d \{{\bf P}_\alpha\} d\{{\bf R}_\alpha\} d\{{\bf p}_{\alpha}\} d\{{\bf q}_{\alpha}\} {\bf \Gamma} e^{-\beta_{N}H_{N}},
\end{equation}
where ${\bf \Gamma}$ has the following expression
\begin{eqnarray}\label{eqn:gamma}
{\bf \Gamma}&=& \frac{2^{(\mathcal{K}+1)N}}{\hbar^{N}}\mathrm{Tr}_\mathrm{e}\prod_{\alpha}\bigg[({\bf q}_{\alpha}+i{\bf p}_{\alpha})({\bf q}_{\alpha}-i{\bf p}_{\alpha})^\mathrm{T}-{\hbar\over2}\mathcal{I}\bigg]\nonumber \\
&&~~~~\times e^{-{1\over\hbar}\sum_{\alpha}({\bf q}_{\alpha}^\mathrm{T}{\bf q}_{\alpha}+{\bf p}_{\alpha}^\mathrm{T}{\bf p}_{\alpha})}. 
\end{eqnarray}

The derived NRPMD Hamiltonian in Eqn.~\ref{eqn:z_wigner}, which is one of the {\it central} results in this paper, has the following expression
\begin{eqnarray}\label{eqn:nrpmd-ham}
&&{H}_{N}= \sum_{\alpha=1}^{N}\bigg[{{{\bf P_{\alpha}}^{2}}\over{2M}}+V_{0}({\bf R_{\alpha}})+{{\bf M}\over{2\beta^{2}_{N}\hbar^{2}}}({\bf R}_{\alpha}-{\bf R}_{\alpha-1})^{2}  \\
&&+ \frac{1}{2\hbar}\sum_{nm}V_{nm}({\bf R}_{\alpha})\big([{\bf q}_{\alpha}]_{n} [{\bf q}_{\alpha}]_{m} + [{\bf p}_{\alpha}]_{n} [{\bf p}_{\alpha}]_{m}-\delta_{nm}\hbar\big)\bigg].\nonumber
\end{eqnarray}
The above Hamiltonian has been proposed in the original NRPMD approach,\cite{richardson2013,richardson2017} and the electronic part of this Hamiltonian ({\it i.e.}, the second line of Eqn.~\ref{eqn:nrpmd-ham}) has also been rigorously derived through a mapping Liouvillian in the state-dependent generalized Kubo-transformed time-correlation function formalism.\cite{hele2016} Here, we provide a rigorous derivation of the full Hamiltonian.

In the NRPMD approach,\cite{richardson2013,richardson2017} classical trajectories are propagated according to the Hamilton's equation of motion associated with ${H}_{N}$ in Eqn.~\ref{eqn:nrpmd-ham}. The motion of the nuclei is governed by $\dot{\bf P}_{\alpha}=-\nabla_{{\bf R}_{\alpha}}H_{N}$ as follows
\begin{eqnarray}\label{eqn:nucforce}
&&\dot{\bf P}_{\alpha}=-{{\bf M}\over{\beta^{2}_{N}\hbar^{2}}}\big(2{\bf R}_{\alpha}-{\bf R}_{\alpha+1}-{\bf R}_{\alpha-1}\big)-\nabla_{{\bf R}_{\alpha}}V_0 ({\bf R}_{\alpha})\\
&&~~-\frac{1}{2\hbar}\sum_{nm}\nabla_{{\bf R}_{\alpha}}V_{nm}({\bf R}_{\alpha})\big([{\bf q}_{\alpha}]_{n} [{\bf q}_{\alpha}]_{m} + [{\bf p}_{\alpha}]_{n} [{\bf p}_{\alpha}]_{m}-\delta_{nm}\hbar\big),\nonumber
\end{eqnarray}
whereas the bead-specific mapping variables are propagated based on the following Hamilton's equation of motion
\begin{eqnarray}\label{eqn:mapeqn}
&&{[\dot{\bf q}_{\alpha}]_{n}=\frac{\partial H_{N}}{\partial [{\bf p}_{\alpha}]_{n}}= \frac{1}{\hbar}\sum_{m}V_{nm}({\bf R}_{\alpha})[{\bf p}_{\alpha}]_{m}}\\
&&{[\dot{\bf p}_{\alpha}]_{n}=-\frac{\partial H_{N}}{\partial [{\bf q}_{\alpha}]_{n}}=-\frac{1}{\hbar}\sum_{m}V_{nm}({\bf R}_{\alpha})[{\bf q}_{\alpha}]_{m}}.\nonumber
\end{eqnarray}

The above NRPMD equations of motion\cite{richardson2013,richardson2017} has been used to compute approximate Kubo-transformed time-correlation functions under thermal conditions, such as the position auto-correlation function,\cite{richardson2013} population auto-correlation function,\cite{richardson2013} and absorption spectra\cite{richardson2017} based on the Fourier transformed dipole auto-correlation function.

\subsection{Excited States Non-adiabatic Dynamics with NRPMD}
With the derived NRPMD Hamiltonian, we propose to investigate the excited states non-adiabatic dynamics. The central quantity that we aim to compute is the following reduced density matrix
\begin{equation}\label{eqn:quantum_corr}
{\rho}_{jj}(t) =\mathrm{Tr}_\mathrm{n}[\hat{\rho}_0e^{i\hat H t/\hbar} \mathcal{\hat P}_j e^{-i\hat H t/\hbar}],
\end{equation}
where, ${\rho}_{jj}(t)$ is the time-dependent population of state $|j\rangle$, $\mathcal{\hat P}_j=|j\rangle\langle j|$ is the projection operator associated with state $|j\rangle$, $\mathrm{Tr}_\mathrm{n}$ represents the trace over the nuclear DOF, and the initial density operator for the entire system is $\hat{\rho}_0=|i\rangle\langle i|\otimes\hat{\rho}_\mathrm{n}$, which is a direct product of the initial electronic state $|i\rangle$ and the initial nuclear density operator $\hat{\rho}_\mathrm{n}$.

Despite that the RPMD-based approaches are originally developed for investigating quantum dynamics under thermal-equilibrium conditions,\cite{habershon2013} recent work\cite{welsch2016} has shown that RPMD yields the exact {\it non-equilibrium} time-correlation function under high temperatures, short-time, and harmonic potential limits through rigorous derivations with the Matsubara dynamics framework\cite{hele2015,hele2015jcp2} which does not subject to any restriction to equilibrium conditions.\cite{welsch2016} In that work, RPMD is used to investigate the non-equilibrium (photoinduced) adiabatic dynamics on a single electronic state, and has shown to accurately describe quantum dynamics compared to the numerically exact results.\cite{welsch2016} Further, the original RPMD method,\cite{menzeelev2010} the MV-RPMD approach\cite{duke2015} and the state-dependent centroid molecular dynamics approach\cite{liao2002} have already been used to simulate non-adiabatic dynamics under {\it non-equilibrium} initial conditions. These early studies inspire us to investigate the numerical performance of NRPMD for simulating photoinduced non-equilibrium dynamics. 

To compute the time-dependent reduced density matrix ${\rho}_{jj}(t)$ in Eqn.~\ref{eqn:quantum_corr} we propose the following NRPMD population expression
\begin{equation}\label{eqn:nrpmd_pop}
{\rho}_{jj}(t)\approx \int d{\boldsymbol \tau} \mathcal{P}^{0}(\{{\bf q}_{\alpha}(0),{\bf p}_{\alpha} (0)\})\rho_\mathrm{rp}(\{{\bf R}_\alpha(0),{\bf P}_\alpha(0)\})\cdot \mathcal{\bar P}_j(t).
\end{equation}
Here, $d{\boldsymbol \tau}\equiv\int d\{{\bf R}_\alpha\}d\{{\bf P}_{\alpha}\}d\{{\bf q}_{\alpha}\}d\{{\bf p}_{\alpha}\}$, with shorthand notation $ d\{{\boldsymbol \chi}_{\alpha}\}= \prod_{\alpha=1}^{N} d{\boldsymbol \chi}_{\alpha}$. In addition, $\mathcal{P}^{0}(\{{\bf q}_{\alpha}(0),{\bf p}_{\alpha}(0)\})$ represents the distribution of the initial electronic mapping variables, $\rho_\mathrm{rp}(\{{\bf R}_\alpha(0),{\bf P}_\alpha(0)\})$ is the ring-polymer density for the initial nuclear density operator $\hat{\rho}_\mathrm{n}$, and $\mathcal{\bar P}_j(t)$ is the time-dependent population estimator. The above proposed expression is akin to the excited states population used in the MV-RPMD approach.\cite{duke2015}

The electronic population estimator $\mathcal{\bar P}$ has many possible choices.\cite{duke2015,richardson2013,richardson2017,hele2016} Here, we use  the following estimator
\begin{equation}\label{eqn:pop_estimator}
\mathcal{\bar P}_j = \frac{1}{N} \sum_{\alpha}\mathcal{P}_j(\alpha) =\frac{1}{N}\sum_{\alpha=1}^{N}{1\over2}([{\bf q}_\alpha]_j^2 +[{\bf p}_\alpha]_j^2 -1), 
\end{equation}
which was originally proposed in the NRPMD approach\cite{richardson2013} and recently derived in various state-dependent RPMD methods,\cite{hele2016,richardson2017} and it is similar to the original MMST population expression.\cite{meyer1979,stock1997}

The initial mapping density $\mathcal{P}^{0}({\bf q}_{\alpha}(0),{\bf p}_{\alpha} (0))$ is required to describe the initial electronic population $\rho_{jj}(0)=\delta_{ij}$, which corresponds to the initial electronic state $|i\rangle$. Here, we adapt the focused initial condition used in the MV-RPMD approach\cite{duke2015,duke2016} and linearized path-integral methods\cite{bonella2003,bonella2005,bonella2008} to represent a constrained mapping density
\begin{equation}
\mathcal{P}^{0}(\{{\bf q}_{\alpha}(0),{\bf p}_{\alpha} (0)\})=\prod_{\alpha=1}^{N}\prod_{j=1}^\mathcal{K}\delta(\mathcal{P}_j(\alpha)-\rho_{jj}(0)).
\end{equation}
The above expression requires the mapping variables to satisfy the following relation
\begin{equation}\label{eqn:estimator}
\mathcal{P}_j (\alpha)= \frac{1}{2}([{\bf q}_\alpha]_j^2 +[{\bf p}_\alpha]_j^2 -1) = \delta_{ij}.
\end{equation}
This can be viewed as the ``Bohr-Sommerfeld" quantization relation\cite{miller2016} that has been used to initialize mapping variables.\cite{meyer1979,muller1999,miller2009,Miller:2016b} Solving the above equation provides the values of the action variables $[{\bf q}_\alpha]_j^2 +[{\bf p}_\alpha]_j^2$ with a value 3 for the occupied state $|i\rangle$, or 1 for unoccupied other states $|j\rangle$, whereas the angle variables $[{\boldsymbol \theta}_{\alpha}]_j=-\tan^{-1}([{\bf p}_{\alpha}]_j/[{\bf q}_{\alpha}]_j)$ are randomly sampled\cite{duke2015,Miller:2016b} within the range of $[0,2\pi]$. Other choices, such as Window estimators,\cite{cotton2016,cotton2013_jcp,Miller:2016b,cotton2019} Wigner transformed projection operators,\cite{richardson2019} or MV-RPMD based estimators\cite{ananth2013,duke2015,duke2016,ananth2017} are possible and subject to future investigations.

To compute non-adiabatic dynamics, real-time trajectories are propagated based on Eqn.~\ref{eqn:nucforce}-\ref{eqn:mapeqn}, with the mapping and nuclear initial distributions sampled from $\mathcal{P}^{0}(\{{\bf q}_{\alpha}(0),{\bf p}_{\alpha} (0)\})$ and $\rho_\mathrm{rp}(\{{\bf R}_\alpha(0),{\bf P}_\alpha(0)\})$, respectively. The time-dependent population is computed from the ensemble average described in Eqn.~\ref{eqn:nrpmd_pop}.


\subsection{Computational Details}
\begin{figure}
 \centering
  \begin{minipage}[t]{1.0\linewidth}
     \centering
     \includegraphics[width=\linewidth]{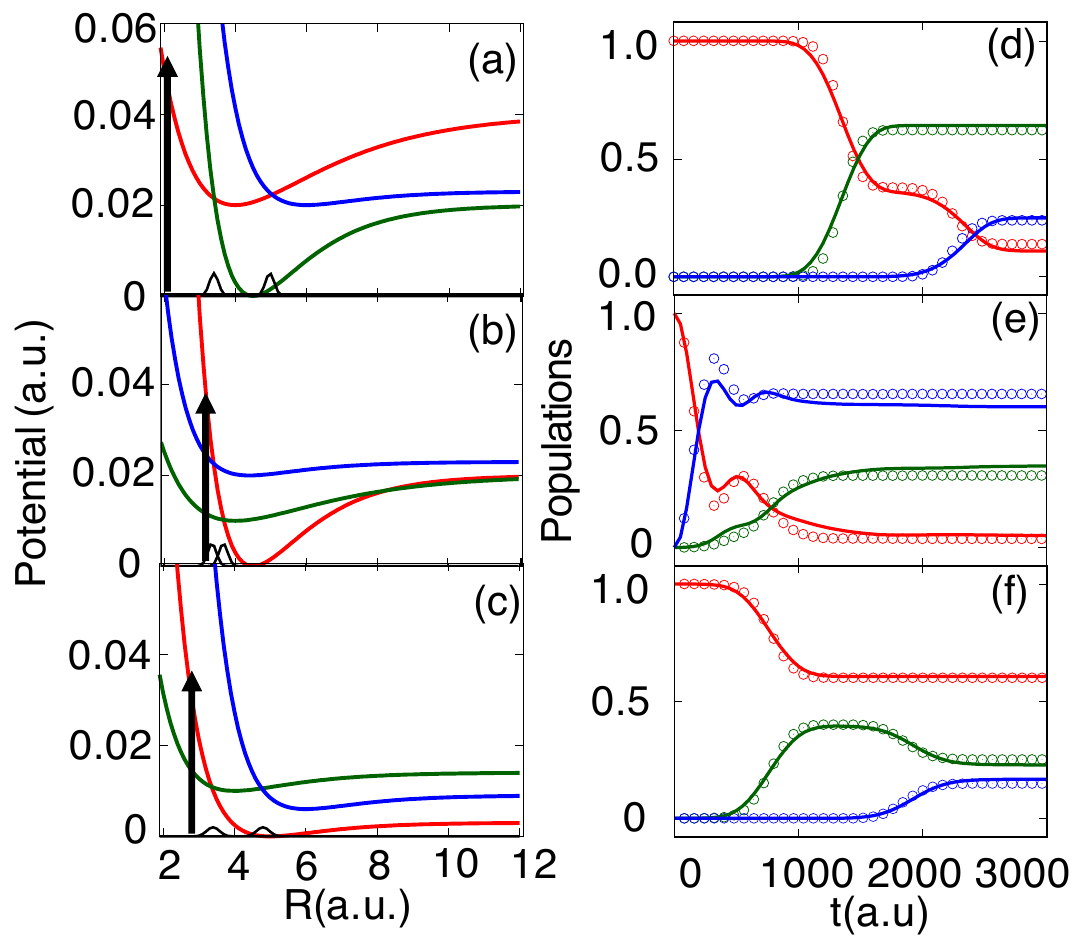}
  \end{minipage}%
   \caption{(a)-(c) present diabatic potentials for Model IA-IC, with diabatic state $|1\rangle$ (red), $|2\rangle$ (green), and $|3\rangle$ (blue). The diabatic couplings are indicated with black solid lines, and the initial Franck-Condon excitation is illustrated with black arrows. The ground electronic state is not shown. (d)-(f) present the real-time population dynamics obtained from the NRPMD propagation (open circles) and numerically exact results (solid lines). The populations are color-coded corresponding to the diabatic surfaces.}
\label{fig:morsecase}
\end{figure}

{\bf Model Systems.} In this paper, we adapt two widely used model systems to investigate the performance of NRPMD for simulating excited state non-adiabatic dynamics. Model I is a widely used three-state Morse potential for photo-dissociation dynamics.\cite{coronado2001} The Hamiltonian operator $\hat{H}=\hat{P}^2/2M+\hat{V}$ of Model I has the following potential
\begin{eqnarray}
V_{ii} &=& \langle i| \hat{V}|i\rangle= D_{ii}(1-e^{-\alpha_{ii}({R}-R_{ii})})^{2} + c_{ii}\\
V_{ij} &=& \langle i| \hat{V}|j\rangle= A_{ij}e^{-\alpha_{ij}({R}-R_{ij})^{2}}.\nonumber
\end{eqnarray}
Here, $V_{ii}$ and $V_{ij}$ are diabatic potentials and couplings, respectively. Parameters of Model I are provided in Table~\ref{tbl:morse}. These potentials and couplings are visualized in Fig.~\ref{fig:morsecase}(a)-(c). The nuclear mass is $M=20,000$ a.u.

Model II is a one dimensional spin-boson system\cite{sato2018,richardson2013,ananth2013} with the following Hamiltonian 
\begin{equation}\label{eqn:1dspinmodel}
\hat{H} = {\hat{P}^{2}\over{2M}}+{1\over2}M\omega^{2}\hat{R}^{2}+\gamma\hat{\sigma_{z}}\otimes \hat{R}+\frac{\Delta}{2}\hat{\sigma}_x.
\end{equation}
In the above expression, $\hat{\sigma}_{z}$ and $\hat{\sigma}_{x}$ are Pauli spin matrices in the electronic subspace $\{|1\rangle, |2\rangle\}$, $\Delta=1$ a.u., $\hat{R}$ and $\hat{P}$ are the position and momentum operators of a harmonic boson mode with frequency $\omega=1$ a.u., and the nuclear mass is $M$=1 a.u. The two-level system and the boson mode interact with each other through a bi-linear coupling potential with a constant coupling strength $\gamma$.
\begin{table} 
\begin{tabular}{ c|ccc|ccc|ccc }
\hline
\hline
 & \multicolumn{3}{c}{Model IA} & \multicolumn{3}{c}{Model IB} & \multicolumn{3}{c}{Model IC} \\
 \hline
$i$&1&2&3&1&2&3&1&2&3\\
 \hline
$D_{ii}$&0.02 &0.02 &0.003 &0.02 &0.01 & 0.003 &0.003 &0.004 &0.003 \\
$\alpha_{ii}$&0.4 &0.65 &0.65 &0.65 &0.4 & 0.65 &0.65 &0.6 &0.65 \\
$R_{ii}$&4.0 &4.5 &6.0 &4.5 & 4.0 & 4.4 & 5.0 & 4.0 & 6.0 \\
$c_{ii}$&0.02 &0.0 & 0.02 & 0.0 & 0.01 & 0.02 & 0.0 & 0.01 & 0.006 \\
  \hline
    \hline
 $ij$ & 12 & 13 & 23 & 12 & 13 & 23& 12 & 13 & 23\\
   \hline
 $A_{ij}$& 0.005 & 0.005 & 0.0 & 0.005 & 0.005 & 0.0 & 0.002 & 0.0 & 0.002 \\
 $\alpha_{ij}$ & 32.0 & 32.0 & 0.0 & 32.0 & 32.0 & 0.0 & 16.0 & 0.0 & 16.0 \\
 $R_{ij}$ & 3.40 & 4.97 & 0.0 & 3.66 & 3.34 & 0.0 & 3.40 & 0.0 & 4.8 \\
  \hline 
  \hline
\end{tabular}
 \caption{Parameters for Model IA-IC (in atomic units).}
 \label{tbl:morse}
\end{table}

{\bf Initial Conditions.} For all results presented in this work, the initial photo-excitation is modeled by the following density operator 
\begin{equation}
\hat{\rho}= |1\rangle\langle1| \otimes \hat{\rho}_\mathrm{n},
\end{equation}
where $|1\rangle$ is the first diabatic state in both models, and the initial nuclear density operator is 
\begin{equation}
\hat{\rho}_\mathrm{n}=e^{-\beta\hat{H}_\mathrm{g}(\hat{R},\hat{P})}. 
\end{equation}
The above initial nuclear density is chosen as the canonical thermal density associated with the ground state Hamiltonian 
\begin{equation}\label{eq:ground}
\hat{H}_\mathrm{g}={\hat{P}^{2}\over{2M}}+{1\over2}M\omega_0^2(\hat{R}-R_{0})^{2}, 
\end{equation}
where $M$ is the nuclear mass, and $R_0$ is the position of the Franck-Condon vertical excitation. Here, we assume that the ground state and the excited states are electronically decoupled; there is no communication between them except during the initial Franck-Condon excitation. For Model I, the initial excitation is indicated with black arrows in Fig.~\ref{fig:morsecase}a-c, with $R_0$=2.1, 3.3, 2.9 for Model IA, IB, and IC, respectively. The frequency of the ground state is chosen\cite{duke2015} as $\omega_0=0.005$ a.u., and $\beta$ is the inverse temperature that corresponds to 300 K. For Model II, $R_{0}=0$, the frequency of the ground state is $\omega_{0}=1$ a.u., and the inverse temperature is $\beta=16$ a.u., such that the initial nuclear quantum distribution is significantly different compared to the classical one, and the dynamics cannot be treated accurately through the classical Wigner model (that samples the initial Wigner distribution and propagate the trajectories classically).

{\bf Simulation Details.} For all NRPMD results presented in this paper, a total of $10^{4}$ trajectories are used to generate the converged population dynamics. The initial thermal nuclear density $\hat{\rho}_n$ is sampled by the normal-mode path-integral Monte-Carlo (PIMC)\cite{tuckerman:1993} in the ground state $\hat{H}_\mathrm{g}$ (Eqn.~\ref{eq:ground}), which generates the ring-polymer initial density 
\begin{equation}\label{eqn:rpinit}
\rho_\mathrm{rp}({\bf R}_\alpha,{\bf P}_\alpha)=e^{-\beta_{N}H^\mathrm{g}_{N}({\bf R}_\alpha, {\bf P}_{\alpha})}.
\end{equation}
In the above equation, $\beta_{N}=\beta/N$ with $N$ as the number of beads (imaginary-time slices), and the ring-polymer Hamiltonian associated with the ground state $\hat{H}_\mathrm{g}$ is expressed as follows 
\begin{equation}
H^\mathrm{g}_{N}=\sum_{\alpha=1}^{N}\frac{{P}_{\alpha}^2}{2M} + \frac{M}{2\beta_{N}^{2}\hbar^2}({R}_\alpha - {R}_{\alpha +1})^2+\frac{1}{2}M\omega_0^2({R}_\alpha - R_{0})^2.
\end{equation}

In this study, we follow the recent works of state-dependent RPMD that treat $N$ as a convergence parameter.\cite{richardson2017,duke2015} For the sampling of the nuclear initial condition (Eqn.~\ref{eqn:rpinit}), a large enough $N$ is used to ensure a converged $\rho_\mathrm{rp}({\bf R}_\alpha,{\bf P}_\alpha)$. This requires $N=4$ for Model I and $N=16$ for Model II, and $N$ remains fixed for the dynamics propagation. We have also performed convergence tests for population dynamics with a higher number of beads, suggesting that these choices of $N$ are sufficient to provide converged results. The initial conditions for the mapping variables are sampled based on Eqn.~\ref{eqn:estimator}. Each configuration is then propagated with Eqn.~\ref{eqn:nucforce}-\ref{eqn:mapeqn}. A symplectic integration scheme is used to numerically propagate the dynamics,\cite{kelly2012,church2018} although other simpler scheme\cite{richardson2017} generates the same numerical results for the model calculations studied here.

The real-time NRPMD dynamics governed by $H_{N}$ (Eqn.~\ref{eqn:nrpmd-ham}) requires a fictitious temperature $\beta$ as the parameter of the dynamics.\cite{Pena2014,duke2015,shakib2017} For Model I, we follow the previous MV-RPMD work\cite{duke2015} on choosing $\beta^{-1}$, which is the energy gap between the zero-point energy (ZPE) of the ground state plus the potential energy gap between the lowest excited state and the initially occupied excited state at $R_{0}$. This provides the fictitious temperatures of 15288 K, 9605 K, and 8843 K for Model IA, IB, and IC,  respectively.\cite{duke2015} For Model II, we directly use $\beta=16$ (associated with the initial distribution) during the NRPMD simulation.

Numerically exact results for Model I are obtained from the discrete variable representation (DVR) calculations,\cite{colbert1992} with a grid spacing of 0.009 a.u. in the range of $R\in [0.5, 20]$ a.u. to ensure convergence. For Model II, exact results are directly obtained by computing the reduced density matrix $\rho_{jj}(t)=\mathrm{Tr}_{\mathrm{n}}[\hat{\rho}_{0}e^{i\hat{H}t/\hbar}|j\rangle\langle j|e^{-i\hat{H}t/\hbar}]$ evaluated with the basis $|i\rangle\otimes|n\rangle$, where $|i\rangle$ is the diabatic basis in Model II and $|n\rangle$ is the eigenstate of the harmonic oscillator centered at $R=0$.

\section{Results and Discussions} 
Fig.~\ref{fig:morsecase} presents the population dynamics of Model IA-IC for photoinduced dissociation dynamics. Panels (a)-(c) present the diabatic potentials. Black arrows indicate the Franck-Condon vertical excitations. Panels (d)-(f) provide the diabatic populations with the same color coding used in (a)-(c) for the diabatic states, obtained from NRPMD simulations (open circles) as well as numerically exact results (solid lines). For all three cases, NRPMD provides a reasonable agreement with the exact results for both the short-time relaxation and non-adiabatic branching dynamics, as well as the longer time asymptotic populations. We emphasize that Model I is a challenging test case for many approximate non-adiabatic dynamics approaches\cite{coronado2001,huo2011molphys,duke2015,tao2016} due to its highly anharmonic potential and non-linear diabatic couplings. Nevertheless, NRPMD provides accurate predictions for the key features of these non-adiabatic events associated with multiple curve crossings.

\begin{figure}
 \centering
  \begin{minipage}[t]{1.0\linewidth}
     \centering
     \includegraphics[width=\linewidth]{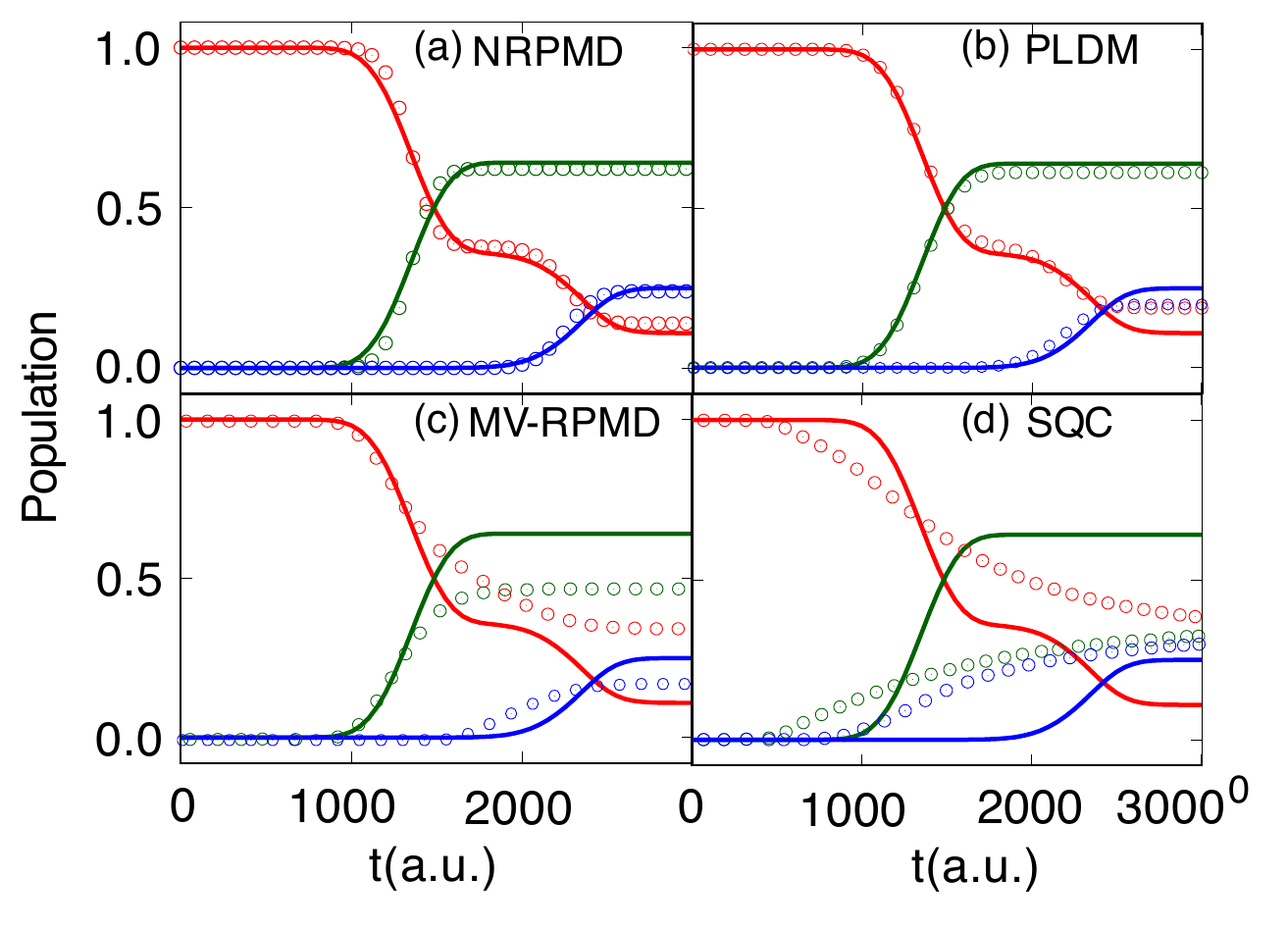}
  \end{minipage}%
   \caption{Population dynamics of Model IA obtaiend from (a) NRPMD, (b) partial linearized density matrix (PLDM) path-integral approach,\cite{huo2011molphys} (c) MV-RPMD,\cite{duke2015} and (d) symmetrical quasi-classical (SQC) approach.\cite{tao2016} The results from these approxiamte trajectory-based approaches are represented by circles, whereas the numerically exact results are depicted with solid lines.}
\label{fig:compare}
\end{figure}
Fig.~\ref{fig:compare} presents the state-dependent population dynamics of Model IA computed from various recently developed dynamics method, including (a) NRPMD (same as the result presented in Fig.~\ref{fig:morsecase}d), (b) partial linearized density matrix (PLDM) path-integral approach,\cite{huo2011molphys,huo2011} (c) MV-RPMD approach,\cite{duke2015,ananth2013} and (d) symmetrical quasi-classical (SQC) approach.\cite{tao2016,miller2016, cotton2013_jcp} In panel (b), PLDM generates accurate short-time non-adiabatic branching dynamics, but starts to deviate from the exact result at a longer time, potentially due to the less accurate partial linearization approximation at a longer time.\cite{huo2011molphys,huo2011} This is confirmed by using an {\it iterative} version of PLDM,\cite{huo2012,lee2016} which only requires linearization approximation for a short time propagator, and then concatenate these short-time PLDM propagators together by important sampling technique.\cite{huo2012,lee2016,bonella2008} The iterative-PLDM provides accurate dynamics for Model I, although a large number of trajectories are required to converge the results.\cite{huo2012} Further, the PLDM Hamiltonian can be viewed as a particular limit of the NRPMD Hamiltonian (or the CS-RPMD Hamiltonian\cite{chowdhury2017}) with one nuclear bead, and two mapping beads for describing the forward and backward propagations.\cite{huo2011} NRPMD which uses multiple beads for all DOFs seems to provide more accurate short-time branching dynamics as well as long-time populations.

In Fig.~\ref{fig:compare}c, MV-RPMD approach\cite{duke2015} provides less accurate non-adiabatic branching dynamics, probably due to the presence of the inter-bead coupling of the mapping variables which contaminates the electronic dynamics.\cite{richardson2013,ananth2013} Further, its inability to correctly capture electronic coherence\cite{althorpe2016} could also mitigate its accuracy of describing the quantum branching dynamics. The NRPMD Hamiltonian, on the other hand, does not contain any inter-bead coupling for mapping variables, thus can reliably capture electronic coherence dynamics\cite{richardson2013,richardson2017} and provide accurate non-adiabatic population transfer as shown in (a).

In Fig.~\ref{fig:compare}d, the recently developed SQC\cite{tao2016,cotton2013_jcp,Miller:2016b} approach provides less accurate results for this model, compared to the other three approaches. Despite that SQC provides accurate non-adiabatic dynamics for many model systems,\cite{Miller:2016b} the population dynamics for Model IA starts to deviate from the exact results even at a very short time. That is also the case for Model IB and IC.\cite{tao2016} Further, the closely related LSC-IVR approach generates very similar results compared to SQC, except some negative populations.\cite{coronado2001,huo2011molphys,duke2015} These results suggest that the Ehrenfest type of the nuclear force, together with the nuclear Wigner distribution (assumed by  both SQC and LSC-IVR) might be the cause for this less accurate dynamics. A recently proposed coherence-controlled SQC (cc-SQC) approach\cite{tao2016} has significantly improved the accuracy by using different nuclear forces based on the time-dependent action variables.\cite{tao2016}

\begin{figure}
 \centering
  \begin{minipage}[t]{0.8\linewidth}
     \centering
     \includegraphics[width=\linewidth]{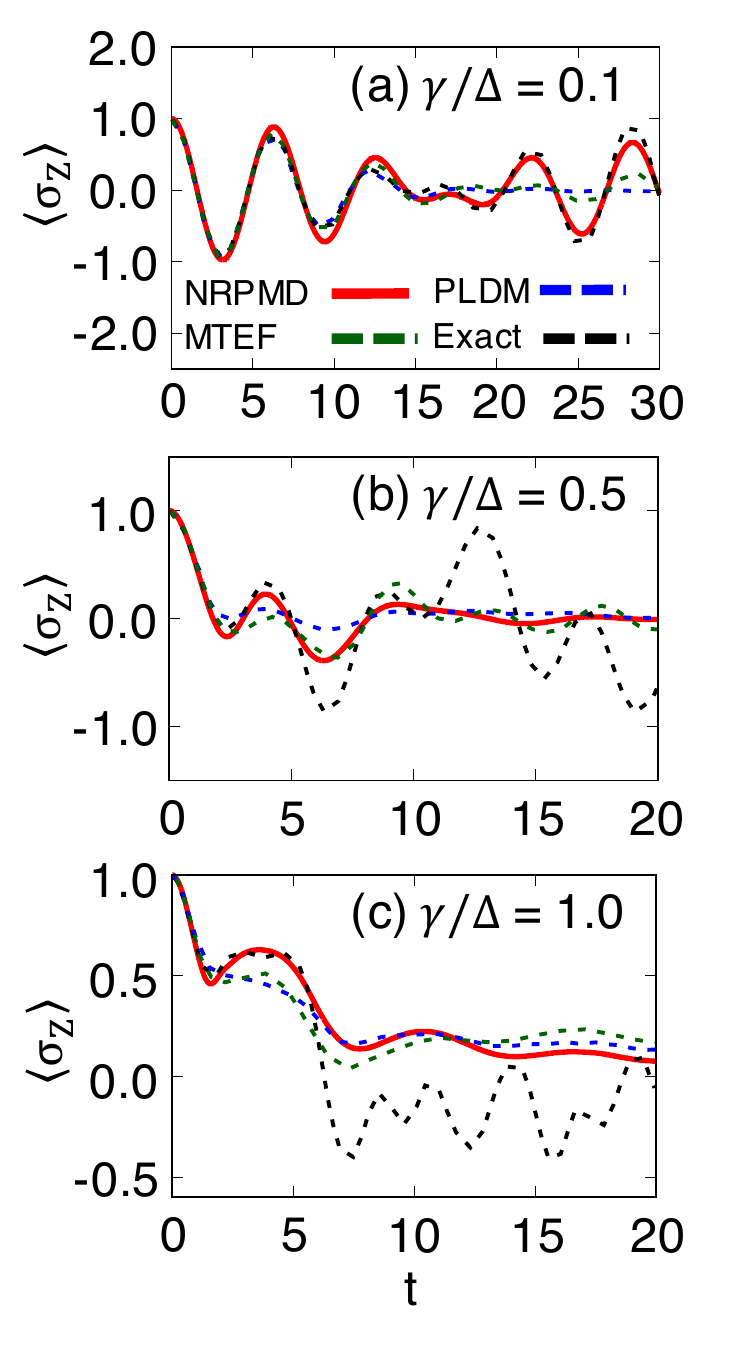}
  \end{minipage}%
   \caption{Population dynamics $\langle\sigma_{z}(t)\rangle$ of Model II with various electron-photon coupling $\gamma/\Delta$. Results are obtained from NRPMD depicted with red solid lines, multitrajectory Ehrenfest dynamics (MTEF) depicted with green dashed line, PLDM depicted with blue dashed lines, and numerically exact results with black dashed lines.}
\label{fig:1dspinboson}
\end{figure}

Fig.~\ref{fig:1dspinboson} presents the population difference $\langle\sigma_{z}(t)\rangle$ between state $|1\rangle$ and $|2\rangle$ in Model II with three different electron-phonon coupling strength $\gamma/\Delta$. Here, we compare the dynamics obtained from NRPMD (red) with multi-trajectory Ehrenfest dynamics (MTEF) (green), PLDM (blue), and numerically exact simulations (black). Panel (a) presents the results in a weak electron-phonon coupling regime with $\gamma/\Delta$=0.1. The temperature $\beta=16$ is low enough such that the initial quantum distribution is significantly different compared to the classical distribution. Thus, quantum mechanical treatment of the nuclear DOF is required for trajectory-based approaches,  through either a Wigner initial distribution (for MTEF and PLDM) or the ring polymer quantization (for NRPMD). One can see that all three approximate methods behave accurately compared to the numerically exact results and reproduce correct oscillations and damping patterns up to $t$ =15 a.u. After that, both the PLDM and MTEF approach fail\cite{sato2018} to accurately describe the longer time recurrence of oscillation in $\langle\sigma_{z}(t)\rangle$. This less accurate longer time dynamics might be caused by the zero point energy (ZPE) leakage problem,\cite{liu2011, habershon2009} which is typical for linearized path-integral approaches based on the classical Wigner dynamics.\cite{Rossky2003,shi2004,miller2009} This ZPE leakage originates from the fact that classical dynamics does not preserve the ZPE incorporated in the initial Wigner distribution,\cite{liu2011,habershon2009} causing an incorrect energy flow from the nuclear DOF to the electronic DOF,\cite{muller1999} equalizing the longer time populations and giving $\langle\sigma_{z}(t)\rangle$=0. Compared to the classical Wigner dynamics of the nuclear DOF,\cite{miller2009,shi2004,Rossky2003} quantizing the nuclear DOF with ring polymer can effectively incorporate nuclear ZPE and describe tunneling effects\cite{shakib2017,shushkov2012,miller2018,miller2019} and alleviate ZPE leaking problem, thus reliably provide the longer time recurrence of the oscillating population.

Fig.~\ref{fig:1dspinboson}b-c presents the population dynamics for stronger electron-phonon couplings. We can see that NRPMD method reproduces the exact result fairly well up to $t$=5 a.u., especially for the model calculation presented in panel (c). At a longer time, however, NRPMD becomes less accurate compared to the exact results, missing the recurrence of the oscillations. These deviations at a longer time in panel (b)-(c) might be due to the intrinsic quantum coherence of nuclear dynamics, which is missed by NRPMD but can be well captured by methods that employ coupled trajectories.\cite{sato2018} Nevertheless, as an independent trajectory-based approach, NRPMD still outperforms both MTEF and PLDM for all model calculations presented here.

We emphasize that the success of any RPMD-based approach relies on the separation of the time-scale between the high-frequency vibrations of the ring polymer and the dynamics of physical interest.\cite{habershon2013} The high-frequency ring polymer oscillations could contaminate the real-time dynamics of the nuclei, which in turn influence the electronic quantum dynamics. This issue could potentially impact the accuracy of our population dynamics, for example, slightly shifting the electronic Rabi oscillation of $\langle\sigma_z (t)\rangle$ in Fig.~\ref{fig:1dspinboson}. To address this issue, we incorporate the recently developed thermostatting technique\cite{ceriotti2010,rossi2014,hele2016_trpmd} to NRPMD, with details provided in Appendix B. Thermostatting has been shown to successfully remove these contaminations from spurious high-frequency oscillations,\cite{hele2016_trpmd} and the model calculations provided in Appendix B indeed confirms that correct electronic Rabi frequency is recovered when applying a Langevin thermostat on the nuclear ring polymer normal modes. Thus, thermostating offers a valuable and practical approach to remove potential contaminations from the high-frequency ring polymer vibrations, enables the promising NRPMD approach to provide more accurate excited states non-adiabatic dynamics.

\section{Conclusion}
In this paper, we present a rigorous derivation of the NRPMD Hamiltonian which was originally proposed in the NRPMD approach.\cite{richardson2013} Our derivation uses the MMST mapping representation for the electronic DOF, and the ring polymer path-integral quantization for the nuclear DOF. The NRPMD Hamiltonian, together with the previously derived CS-RPMD Hamiltonian,\cite{chowdhury2017} can be viewed as a unified theory for classical electronic states (through the MMST mapping description) and classical nuclei (through the ring polymer quantization).

We further propose to compute excited state non-adiabatic quantum dynamics\cite{duke2015,welsch2016} with NRPMD Hamiltonian. Numerical results with the coupled Morse potential and one-dimensional spin-boson model suggest that NRPMD can provide accurate short-time branching dynamics and a reliable longer time dynamics. The NRPMD Hamiltonian does not contain inter-bead coupling terms associated with the electronic mapping variables; thus it is capable to accurately capture the electronic quantum dynamics.\cite{richardson2013,richardson2017} In this particular model calculation, NRPMD outperforms the recently developed MV-RPMD methods\cite{duke2015,ananth2013} which does include these inter-bead coupling terms that might contaminate the electronic dynamics. Compared to the linearized semi-classical methods based on Wigner quantization of nuclei,\cite{miller2009,shi2004,huo2011} quantizing nuclear DOF with ring polymer can effectively incorporate nuclear zero-point energy (ZPE) and alleviates ZPE leakage problem. 

This work opens up new possibilities of using state-dependent RPMD approaches\cite{richardson2013,richardson2017,chowdhury2017} to accurately simulate electronic non-adiabatic dynamics. These approaches are potentially well-suited theoretical methods to investigate photochemical reactions, especially when nuclear quantum effects play an important role, such as in the photoinduced proton-coupled electron transfer reactions.\cite{Venkataraman:2010,Hazra:2010,Hazra:2011,Goyal:2015,Goyal:2016,Goyal:2017,Mandal2018jcp} We note that the original RPMD method is limited to one electron non-adiabatic process,\cite{habershon2013,menzeelev2010,menzeelev2011,kretchmer2013,kretchmer2016,Ananth2016} whereas the vibronic quantization approach (which uses an explicit quantum-state description of proton) can be numerically expensive for a three-dimensional quantum treatment of many protons.\cite{Hammes-Schiffer:2015,Hazra:2010,Hazra:2011,Goyal:2016,Goyal:2017,Mandal2018jcp,Mandal2019jpca} State-dependent RPMD approaches, such as NRPMD\cite{richardson2013} and CS-RPMD,\cite{chowdhury2017} provide accurate electronic non-adiabatic dynamics and nuclear quantum effects without any limitations on the number of electrons and protons that can be explicitly described. 

We note that most of the state-dependent RPMD approaches, including the NRPMD method discussed here, are formulated in the {\it diabatic} representation. To perform on-the-fly simulation with adiabatic electronic structure calculations, these approaches are usually reformulated back to the adiabatic representation. This process requires nontrivial theoretical efforts, and the adiabatic equation of motion are computationally inconvenient due to the presence of derivative couplings. These non-trivial tasks, however, can be avoided by using the recently developed quasi-diabatic propagation scheme.\cite{Mandal:2018} The QD scheme uses the adiabatic states associated with a reference geometry as the local diabatic states during a short-time propagation step, and dynamically update the definition of the diabatic states along the time-dependent nuclear trajectory. This scheme thus allows a seamless interface between diabatic dynamics approaches (such as NRPMD) with adiabatic electronic structure calculations, providing new frameworks to accurately and efficiently perform non-adiabatic on-the-fly simulations.

Future investigations will also focus on analytic derivation of the NRPMD {\it dynamics} based on rigorous theoretical frameworks,\cite{Geva2003,jang14,hele2015,hele2015jcp2} such as the Matsubara dynamics\cite{hele2015,hele2015jcp2} and the exact mapping Liouvillian.\cite{hele2016} These formal theoretical derivations will help to assess the validity and the accuracy of the state-dependent RPMD approaches for simulating excited states non-adiabatic dynamics. 

\section{Acknowledgement} 
This work was supported by the National Science Foundation CAREER Award under Grant No. CHE-1845747. Computing resources were provided by the Center for Integrated Research Computing (CIRC) at the University of Rochester.

\section{Appendix A: Details of the derivation for NRPMD partition function}
In this appendix, we provide details of derivations for (i) the expression of $[{\hat H}_{e}({\bf R}_\alpha)]_{\mathrm{w}_\alpha}$ in Eqn.~\ref{eqn:mapwig}, (ii) the expression of the electronic partition function $\mathrm{Tr_e}[\prod_{\alpha=1}^{N}e^{-\beta_N \hat{H}_e({\bf R}_\alpha)}]$ in Eqn.~\ref{eqn:elec-part}, and (iii) the expression of $\boldsymbol\Gamma$ in Eqn.~\ref{eqn:gamma}.

First, we provide the derivation of the $[{\hat H}_\mathrm{e}({\bf R}_\alpha)]_{\mathrm{w}_{\alpha}}$ expression. The Wigner transformed mapping Hamiltonian operator (Eqn.~\ref{eqn:state_dep_pot}) can be expressed as follows
\begin{widetext}
\begin{eqnarray}\label{eqn:detail_a}
&&[{\hat H}_{e}({\bf R}_\alpha)]_{\mathrm{w}_\alpha}=\int d{\bf \Delta}_{\alpha} e^{i{\bf p}_{\alpha}^{\mathrm{T}}{\bf \Delta}_{\alpha}/\hbar} \bigg\langle {\bf q}_{\alpha}-\frac{{\bf \Delta}_{\alpha}}{2}\bigg | \sum_{nm}V_{nm}({\bf R}_\alpha)\frac{1}{2\hbar}\bigg({\hat q}_{n} {\hat q}_{m} + {\hat p}_{n} {\hat p}_{m}-\delta_{nm}\hbar\bigg)\bigg|{\bf q}_{\alpha}+ \frac{{\bf \Delta}_{\alpha}}{2} \bigg\rangle \\
&&= \frac{1}{2\hbar}\sum_{nm}V_{nm}({\bf R}_\alpha)\int d{\bf \Delta}_{\alpha}e^{i{\bf p}_{\alpha}^{T}{\bf \Delta}_{\alpha}/\hbar}\bigg[ \bigg({\bf q}_{\alpha}-\frac{{\bf \Delta}_{\alpha}}{2}\bigg)_{n}\bigg({\bf q}_{\alpha}+\frac{{\bf \Delta}_{\alpha}}{2}\bigg)_{m} -\hbar^{2}\frac{\partial}{\partial [{\bf \Delta}_{\alpha}]_{n}}\frac{\partial}{\partial [{\bf \Delta}_{\alpha}]_{m}}-\delta_{nm}\hbar \bigg] \delta(\bf \Delta_{\alpha})\nonumber\\
&&= \frac{1}{2\hbar}\sum_{nm}V_{nm}({\bf R}_\alpha)\bigg[[{\bf q}_{\alpha}]_{n}[{\bf q}_{\alpha}]_{m}-\hbar^{2}\int d{\bf \Delta}_{\alpha} \delta({\bf \Delta}_{\alpha})\frac{\partial}{\partial [{\bf \Delta}_{\alpha}]_{n}}\frac{\partial}{\partial [{\bf \Delta}_{\alpha}]_{m}}e^{i{\bf p}_{\alpha}^{\mathrm{T}}{\bf \Delta}_{\alpha}/\hbar}-\delta_{nm}\hbar\int d{\bf \Delta}_{\alpha}\delta({\bf \Delta}_{\alpha})e^{i{\bf p}_{\alpha}^\mathrm{T}{\bf \Delta}_{\alpha}}\bigg]\nonumber\\
&&=\frac{1}{2\hbar}\sum_{nm}V_{nm}({\bf R}_\alpha)\bigg([{\bf q}_{\alpha}]_{n}[{\bf q}_{\alpha}]_{m}+ [{\bf p}_{\alpha}]_{n}[{\bf p}_{\alpha}]_{m}-\delta_{nm}\hbar\bigg).\nonumber
\end{eqnarray}
\end{widetext}
Between the first and the second equality, we have used the fact $\langle q-\frac{\Delta}{2}|\hat{p}|q+\frac{\Delta}{2}\rangle=\int dp p \langle q-\frac{\Delta}{2}|p\rangle\langle p|q+\frac{\Delta}{2}\rangle= (\frac{1}{2\pi\hbar})\int dp p e^{-\frac{i}{\hbar}p\Delta}= (\frac{1}{2\pi\hbar})i\hbar\frac{\partial}{\partial \Delta}\int dp e^{-\frac{i}{\hbar}p\Delta}=i\hbar\frac{\partial}{\partial \Delta}\int dp \langle q-\frac{\Delta}{2}|p\rangle\langle p|q+\frac{\Delta}{2}\rangle=i\hbar\frac{\partial}{\partial \Delta}\langle q-\frac{\Delta}{2}|q+\frac{\Delta}{2}\rangle$. Between the second and the third equality, we use integration by parts for $\int d{\bf \Delta}_{\alpha}$.  Analytically perform the rest integrals, we arrived at the final expression of $[{\hat H}_{e}({\bf R}_\alpha)]_{\mathrm{w}_\alpha}$ in Eqn.~\ref{eqn:mapwig}. A similar derivation has been provided in the previous work by Kapral and co-workers\cite{kapral2010} for developing the Wigner mapping mixed quantum-classical Liouville (MQCL) equation. 

Second, we provide the details of the derivation for the electronic trace expression in Eqn.~\ref{eqn:elec-part}. We start by inserting a resolution of identity $\int d {\bf q}_{2} \mathcal {P} |{\bf q}_2\rangle\langle {\bf q}_{2}|$ into $\left[\prod_{\alpha = 2}^{N} e^{-\beta_{N}\hat{H}_{e}({\bf R}_\alpha)}\mathcal {P}\right]_{\mathrm{w}_1}$, resulting in the second line of Eqn.~\ref{eqn:trace-wig}. Next, we alter the order of terms in line 3-4 of Eqn.~\ref{eqn:trace-wig}, leading to an integral over ${\bf q}_{2}$. Finally, replacing $\int d{\bf q}_{2}$ by the corresponding Wigner transform,\cite{duke2015} we can use the property in Eqn.~\ref{eqn:wig-prop} again and repeat the same procedure outlined in Eqn.~\ref{eqn:power_series} to factorize the total Wigner function as a product of two Wigner functions inside the $\int d{\bf q}_2 d{\bf p}_2$ integral, arriving at the last line of Eqn.~\ref{eqn:trace-wig}.
\begin{widetext}
\begin{eqnarray}\label{eqn:trace-wig}
&&\text{Tr}_\mathrm{e}\left[\prod_{\alpha = 1}^{N} e^{-\beta_{N} \hat{H}_{e}({\bf R}_\alpha)}\right]\\
&=&\frac{1}{(2\pi\hbar)^\mathcal{K}}\int d{\bf q}_{1}d{\bf p}_{1} e^{-\beta_N [\hat H_{e}({\bf R}_{1})]_\mathrm{w_1}}\left[\int d {\bf q}_{2} \mathcal {P} |{\bf q}_2\rangle\langle {\bf q}_{2}|\prod_{\alpha = 2}^{N} e^{-\beta_{N}\hat{H}_{e}({\bf R}_\alpha)}\mathcal {P}\right]_\mathrm{w_1} \nonumber\\
&=&\frac{1}{(2\pi\hbar)^\mathcal{K}}\int d{\bf q}_{1}d{\bf p}_{1} e^{-\beta_N [\hat H_{e}({\bf R}_{1})]_\mathrm{w_1}} \int d{\bf \Delta}_{1}e^{i{\bf p}^\mathrm{T}{\bf \Delta}_{1}/\hbar} \big\langle {\bf q}_{1}-\frac{{\bf \Delta}_1}{2}\big| \int d {\bf q}_{2} \mathcal {P} |{\bf q}_2\rangle\langle {\bf q}_{2}| \prod_{\alpha = 2}^{N} e^{-\beta_{N}\hat{H}_{e}({\bf R}_\alpha)}\mathcal {P}\big|{\bf q}_{1}+\frac{{\bf \Delta}_1}{2}\big\rangle\nonumber\\
&=&\frac{1}{(2\pi\hbar)^\mathcal{K}}  \int d{\bf q}_{1}d{\bf p}_{1} e^{-\beta_N [\hat H_{e}({\bf R}_{1})]_\mathrm{w_1}}\int d{\bf \Delta}_{1}  e^{i{\bf p}^\mathrm{T}{\bf \Delta}_{1}/\hbar}\int d {\bf q}_{2}\big\langle {\bf q}_{2} \big|e^{-\beta_{N}\hat{H}_{e}({\bf R}_2)} \prod_{\alpha = 3}^{N} e^{-\beta_{N}\hat{H}_{e}({\bf R}_\alpha)}\mathcal {P}\big|{\bf q}_{1}+\frac{{\bf \Delta}_1}{2}\rangle\langle {\bf q}_{1}-\frac{{\bf \Delta}_1}{2}|\mathcal {P}\big|{\bf q}_{2}\big\rangle \nonumber\\
&=&\frac{1}{(2\pi\hbar)^{2\mathcal{K}}} \int d{\bf q}_{1}d{\bf p}_{1} e^{-\beta_N [\hat H_{e}({\bf R}_{1})]_\mathrm{w_1}} \int d{\bf \Delta}_{1}  e^{i{\bf p}_{1}^\mathrm{T}{\bf \Delta}_{1}/\hbar} \int d {\bf q}_{2}d {\bf p}_{2} \big[e^{-\beta_{N}\hat{H}_{e}({\bf R}_2)} \prod_{\alpha = 3}^{N} e^{-\beta_{N}\hat{H}_{e}({\bf R}_\alpha)}\mathcal {P}|{\bf q}_{1}+\frac{{\bf \Delta}_1}{2}\rangle\langle {\bf q}_{1}-\frac{{\bf \Delta}_1}{2}| \mathcal{P} \big]_\mathrm{w_2}\nonumber\\
&=&\frac{1}{(2\pi\hbar)^{2\mathcal{K}}} \int d{\bf q}_{1}d{\bf p}_{1} e^{-\beta_N [\hat H_{e}({\bf R}_{1})]_\mathrm{w_1}}\int d{\bf \Delta}_{1}  e^{i{\bf p}_{1}^\mathrm{T}{\bf \Delta}_{1}/\hbar} \int d {\bf q}_{2}d {\bf p}_{2} e^{-\beta_{N}[\hat{H}_{e}({\bf R}_2)]_\mathrm{w_2}} \bigg[\prod_{\alpha = 3}^{N} e^{-\beta_{N}\hat{H}_{e}({\bf R}_\alpha)}\mathcal {P}|{\bf q}_{1}+\frac{{\bf \Delta}_1}{2}\rangle\langle {\bf q}_{1}-\frac{{\bf \Delta}_1}{2} |\mathcal {P}\bigg]_\mathrm{w_2}\nonumber
\end{eqnarray}
\end{widetext}
A similar procedure of the above derivation has been recently used to derive the population estimator in the MV-RPMD approach.\cite{duke2015}

Third, we derive the expression of $\boldsymbol\Gamma$ (Eqn.~\ref{eqn:gamma}). We start from the last line of Eqn.~\ref{eqn:elec-part} and define 
\begin{equation}\label{eqn:detail_electronic_part}
\boldsymbol\Gamma = \prod_{\alpha =1}^N \int d\boldsymbol\Delta_{\alpha}e^{i{\bf p}_{\alpha}^{\mathrm{T}}\boldsymbol\Delta_{\alpha}/\hbar}\langle {\bf q}_{\alpha} - \frac{\boldsymbol\Delta_{\alpha}}{2} |\mathcal{P}|{\bf q}_{\alpha+1} + \frac{\boldsymbol\Delta_{\alpha+1}}{2}\rangle
\end{equation}
where $\mathcal{P}= \sum_{n}|n\rangle\langle n|$ is the projection operator in the SEO basis. Recall that the SEO mapping wavefunction is the product of $(\mathcal{K}-1)$ ground state harmonic oscillator wavefunctions and one excited state harmonic oscillator wavefunction
\begin{equation}\label{eqn:SEO_wavefunction}
\langle {\bf q}|n \rangle = \sqrt{\frac{2}{\hbar}}\frac{1}{(\pi\hbar)^{\mathcal{K}/4}}[{\bf q}]_{n}e^{-{\bf q}^{\mathrm{T}}{\bf q}/2\hbar}.
\end{equation}
With the above relation, we can rewrite Eqn.~\ref{eqn:detail_electronic_part} as 
\begin{eqnarray}\label{eqn:gaussian_overlap}
\boldsymbol{\Gamma} &=& \frac{2^N}{\hbar^N}\frac{1}{(\pi\hbar)^{N\mathcal{K}/{2}}}\prod_{\alpha=1}^{N}\int d\boldsymbol\Delta_{\alpha}\big({\bf q}_{\alpha}-\frac{\boldsymbol\Delta_{\alpha}}{2}\big)^{\mathrm{T}} \big({\bf q}_{\alpha +1}+\frac{\boldsymbol\Delta_{\alpha +1}}{2}\big)\nonumber\\
&&~~~~\times e^{-\frac{1}{\hbar}(\frac{1}{4}\boldsymbol\Delta_{\alpha}^{\mathrm{T}}\boldsymbol\Delta_{\alpha}+{\bf q}_{\alpha}^{\mathrm{T}}{\bf q}_{\alpha}-i{\bf p}_{\alpha}^{\mathrm{T}}\boldsymbol\Delta_{\alpha})}
\end{eqnarray}
Rearranging the prefactors of the above equation and grouping terms associated with $\boldsymbol\Delta_{\alpha}$, we have
\begin{eqnarray}\label{eqn:group_delta}
\boldsymbol{\Gamma}&=&\frac{2^N}{\hbar^N}\frac{1}{(\pi\hbar)^{N\mathcal{K}/2}} \int d\{{\bf \Delta_{\alpha}}\}\\
&&~~~\times \mathrm{Tr}_\mathrm{e}\big[\prod_{\alpha=1}^{N}\big({\bf q}_{\alpha}+\frac{\boldsymbol\Delta_{\alpha}}{2}\big)\otimes \big({\bf q}_{\alpha}-\frac{\boldsymbol\Delta_{\alpha}}{2}\big)^{\mathrm{T}}\big]\nonumber\\
&&~~~\times e^{-\frac{1}{\hbar}\sum_{\alpha=1}^{N}(\frac{1}{4}\boldsymbol\Delta_{\alpha}^{\mathrm{T}}\boldsymbol\Delta_{\alpha}+{\bf q}_{\alpha}^{\mathrm{T}}{\bf q}_{\alpha}-i{\bf p}_{\alpha}^{\mathrm{T}}\boldsymbol\Delta_{\alpha})}\nonumber
\end{eqnarray}
Analytically performing the integration over $\boldsymbol\Delta_{\alpha}$ (a Gaussian integral), we obtain the final expression in Eqn.~\ref{eqn:gamma}. Similar derivations can also be found in the previous work of MV-RPMD\cite{ananth2013} as well as in the recently derived exact mapping variable Liouvillian.\cite{hele2016}

\section{Appendix B: Thermosttated NRPMD (T-NRPMD)}
In this appendix, we investigate the effect of thermostatting nuclear ring polymer on the NRPMD non-adiabatic dynamics. It is known that ring polymer quantization often introduces spurious frequencies in RPMD dynamics due to the presence of the high-frequency normal mode vibrations,\cite{habershon2013} and causes the ``spurious resonance problem'' for computing spectra\cite{ceriotti2010,rossi2014,hele2016_trpmd} and introducing incorrect frequency in time-correlation functions for nonlinear operators.\cite{hele2016_trpmd} Thus, the success of any RPMD-type approach relies on the separation of the time-scale between the high-frequency normal mode vibrations of the ring polymer and the dynamics of physical interest.\cite{habershon2013} Various thermostatted RPMD (TRPMD) approaches\cite{ceriotti2010,rossi2014} are proposed to achieve this.\cite{ceriotti2010,rossi2014}

Based on the Matsubara dynamics framework, it is recently shown that this frequency contamination arises due to discarding the imaginary term of the Matsubara Liouvillian\cite{hele2015,hele2015jcp2} when deriving the RPMD approach. This formal analysis\cite{hele2016_trpmd} shows that TRPMD can be justified by {\it replacing} the imaginary Matsubara Liouvillian with a friction term, such as the Fokker-Planck operator, \cite{bussi2007,gardiner2003,risken1989} instead of just discarding the imaginary part of the Matsubara Liouvillian as done in RPMD. Here, we apply a Langevin thermostat that couples to the nuclear normal mode in NRPMD. We briefly introduce the normal mode representation of the ring polymer, before we provide the equation of motion for thermostatting.

The free ring-polymer Hamiltonian (see Eqn.~\ref{eqn:hrp} and below) is defined as follows
\begin{equation}\label{eqn:freerp}
H_\mathrm{rp}=\sum_{\alpha=1}^{N}{{{\bf P_{\alpha}}^{2}}\over{2{\bf M}}}+{{\bf M}\over{2\beta^{2}_{N}\hbar^{2}}}({\bf R_{\alpha}-R_{\alpha-1}})^{2}.
\end{equation}
Often, the dynamical propagation of RPMD (and PIMD) can be simplified by transforming $H_\mathrm{rp}$ from the above bead representation (or so-called the primitive nuclear coordinate) to the normal mode representation, which is the eigenstate of the Hessian matrix of $H_\mathrm{rp}$. Diagonalizing the Hessian matrix of $H_\mathrm{rp}$ provides the eigenvalue, {\it i.e.}, the following normal mode frequency
\begin{equation}\label{eqn:normfreq}
\widetilde{\omega}_{\mu}=\frac{2}{\beta_{N}\hbar}\sin\left(\frac{\mu\pi}{N}\right),
\end{equation}
where $\mu\in[0,N-1]$ represents the index of the normal mode. The same diagonalization process also gives the eigenvector $T_{\alpha\mu}$ of the Hessian matrix, which provides the relation between the primitive coordinate $\{{\bf R}_{\alpha}\}$ and the normal mode coordinate $\{\widetilde{\bf R}_{\mu}\}$, as well as the corresponding relation for momenta under two representations. These relations are expressed as follows
\begin{eqnarray}
&&\widetilde{\bf R}_{\mu}=\sum_{\alpha=1}^{N}{\bf R}_{\alpha}T_{\alpha\mu};~~~~\widetilde{\bf P}_{\mu}=\sum_{\alpha=1}^{N}{\bf P}_{\alpha}T_{\alpha\mu};\label{eqn:trans}\\
&&{\bf R}_{\alpha}=\sum_{\mu=0}^{N-1}T_{\alpha\mu}\widetilde{\bf R}_{\mu};~~~~{\bf P}_{\alpha}=\sum_{\mu=0}^{N-1}T_{\alpha\mu}\widetilde{\bf P}_{\mu}.\label{eqn:backtrans}
\end{eqnarray}
The above transformation matrix elements have the following values
\begin{eqnarray}\label{eqn:transcoeff}
T_{\alpha\mu}=
\begin{cases}
  \sqrt{1/N}~~~~~~~~~~~~~~~~~~~~~~~(\mu=0)\\    
  \sqrt{2/N}\cos(2\pi\alpha\mu/N)~~~~~(1 \le \mu \le \frac{N}{2}-1)\nonumber\\
  \sqrt{1/N}(-1)^{\alpha}~~~~~~~~~~~~~~~(\mu=\frac{N}{2})\\
  \sqrt{2/N}\sin(2\pi\alpha\mu/N)~~~~~~(\frac{N}{2}+1\le \mu \le {N-1}).
\end{cases}\\
\end{eqnarray}
Under the normal mode representation, the free ring polymer Hamiltonian $H_\mathrm{rp}$ in Eqn.~\ref{eqn:freerp} becomes
\begin{equation}
H_\mathrm{rp}=\sum_{\mu=0}^{N-1} {{{\widetilde{\bf P}_{\mu}}^{2}}\over{2{\bf M}}}+\frac{1}{2}{\bf M}\widetilde{\omega}_{\mu}^{2}\widetilde{\bf R}_{\mu}^2,
\end{equation}
where the normal mode frequency $\widetilde{\omega}_{\mu}$ is described in Eqn.~\ref{eqn:normfreq}. Note that the inter-bead coupling terms of the ring polymer become a set of simple quadratic terms with the normal mode frequencies. The nuclear equation of motion of NRPMD described in Eqn.~\ref{eqn:nucforce} under the normal mode representation is expressed as follows
\begin{equation}\label{eqn:nmforce}
{\dot{\widetilde{\bf P}}_{\mu}}=-\nabla_{\widetilde{\bf R}_{\mu}}H_{N}(\{{\bf R}_{\alpha}\})=-\sum_{\alpha}\nabla_{{\bf R}_{\alpha}}H_{N}(\{{\bf R}_{\alpha}\})\frac{\partial {\bf R}_{\alpha}}{\partial\widetilde{\bf R}_{\mu}}
\end{equation}
where simple chain rule is used to establish the last equality, $-\nabla_{{\bf R}_{\alpha}}H_{N}(\{{\bf R}_{\alpha}\})=\dot{\bf P}_{\alpha}$ is the nuclear force in Eqn.~\ref{eqn:nucforce}, and ${T}_{\alpha\mu}={\partial {\bf R}_{\alpha}}/{\partial\widetilde{\bf R}_{\mu}}$ is the Jacobian matrix element of the transformation between the primitive nuclear variables $\{{\bf R}_{\alpha}\}$ and the normal mode coordinates $\{\widetilde{\bf R}_{\mu}\}$ described in Eqn.~\ref{eqn:backtrans}. Note that ${T}_{\alpha\mu}$ is same for all $\mathcal{F}$ nuclear DOF. 

The normal mode NRPMD nuclear force $-\nabla_{\widetilde{\bf R}_{\mu}}H_{N}(\{{\bf R}_{\alpha}\})$ in Eqn.~\ref{eqn:nmforce} contains three types of terms: (i) the force contribution from the free ring polymer, $-{\bf M}\omega_{\mu}^{2}\widetilde{\bf R}_{\mu}$, (ii) the state independent force, $-\sum_{\alpha}\nabla_{{\bf R}_{\alpha}} V_0 ({\bf R}_{\alpha}){T}_{\alpha\mu} $, and (iii) the state dependent force, $-\frac{1}{2\hbar}\sum_{\alpha}\sum_{nm}\nabla_{{\bf R}_{\alpha}}V_{nm}({\bf R}_{\alpha})\big([{\bf q}_{\alpha}]_{n} [{\bf q}_{\alpha}]_{m} + [{\bf p}_{\alpha}]_{n} [{\bf p}_{\alpha}]_{m}-\delta_{nm}\hbar\big){T}_{\alpha\mu}$.

Following the previous work of TRPMD,\cite{ceriotti2010,hele2016_trpmd} the nuclear ring polymer normal mode $\{ \widetilde{\bf R}_{\mu}\}$ in $H_{N}$ is coupled to an Langevin thermostat, giving a method that we referred as the Thermostatted NRPMD (T-NRPMD). In T-NRPMD, the mapping equations of motion remain the same as described in Eqn~\ref{eqn:mapeqn}, whereas the nuclear equation of motion in Eqn.~\ref{eqn:nmforce} is replaced\cite{ceriotti2010,hele2016_trpmd} by the following one 
\begin{equation}\label{trpmd_normal_mode}
{\dot{\widetilde{\bf P}}_{\mu}} = -\nabla_{\widetilde{\bf R}_{\mu}}H_{N}(\{{\bf R}_{\alpha}\})-{\boldsymbol \eta}_{\mu}\widetilde{\bf P}_{\mu} + \sqrt{\frac{2{\bf M}{\boldsymbol\eta}_{\mu}}{\beta_{N}}}{\boldsymbol \xi}_{\mu}(t). 
\end{equation}
The first term is the force from the NRPMD Hamiltonian (Eqn.~\ref{eqn:nmforce}), the second term is the friction force acting on $\widetilde{\bf P}_{\mu}$ with ${\boldsymbol\eta}_{\mu}$ as the bead-specific normal mode friction matrix, and the last term is the random force, with ${\boldsymbol \xi}_{\mu}(t)$ representing an uncorrelated, Gaussian-distributed random force\cite{ceriotti2010} with unit variance
$\langle{\boldsymbol\xi}_{\mu}(0){\boldsymbol \xi}_{\mu}(t)\rangle=\delta(t)$, and zero mean $\langle{\boldsymbol \xi}_{\mu}(t)\rangle=0$. Based on the recent analysis of TRPMD from the Matsubara dynamics framework,\cite{hele2016_trpmd} we choose the same friction constant for all $\mathcal {F}$ nuclear DOF associated with the $\mu_\mathrm{th}$ normal mode, with the $\mu$-specific friction term, $\eta_{\mu} = 2\lambda|\widetilde{\omega}_\mu|$, where $\lambda$ is viewed as a parameter.\cite{hele2016_trpmd} The Langevin equation can be numerically propagated based on the algorithm in previous works,\cite{ceriotti2010,bussi2007} whereas the mapping equation of motion is integrated with a symplectic integrator.\cite{kelly2012,church2018}
\begin{figure}[h!]
 \centering
  \begin{minipage}[t]{1.0\linewidth}
     \centering
     \includegraphics[width=\linewidth]{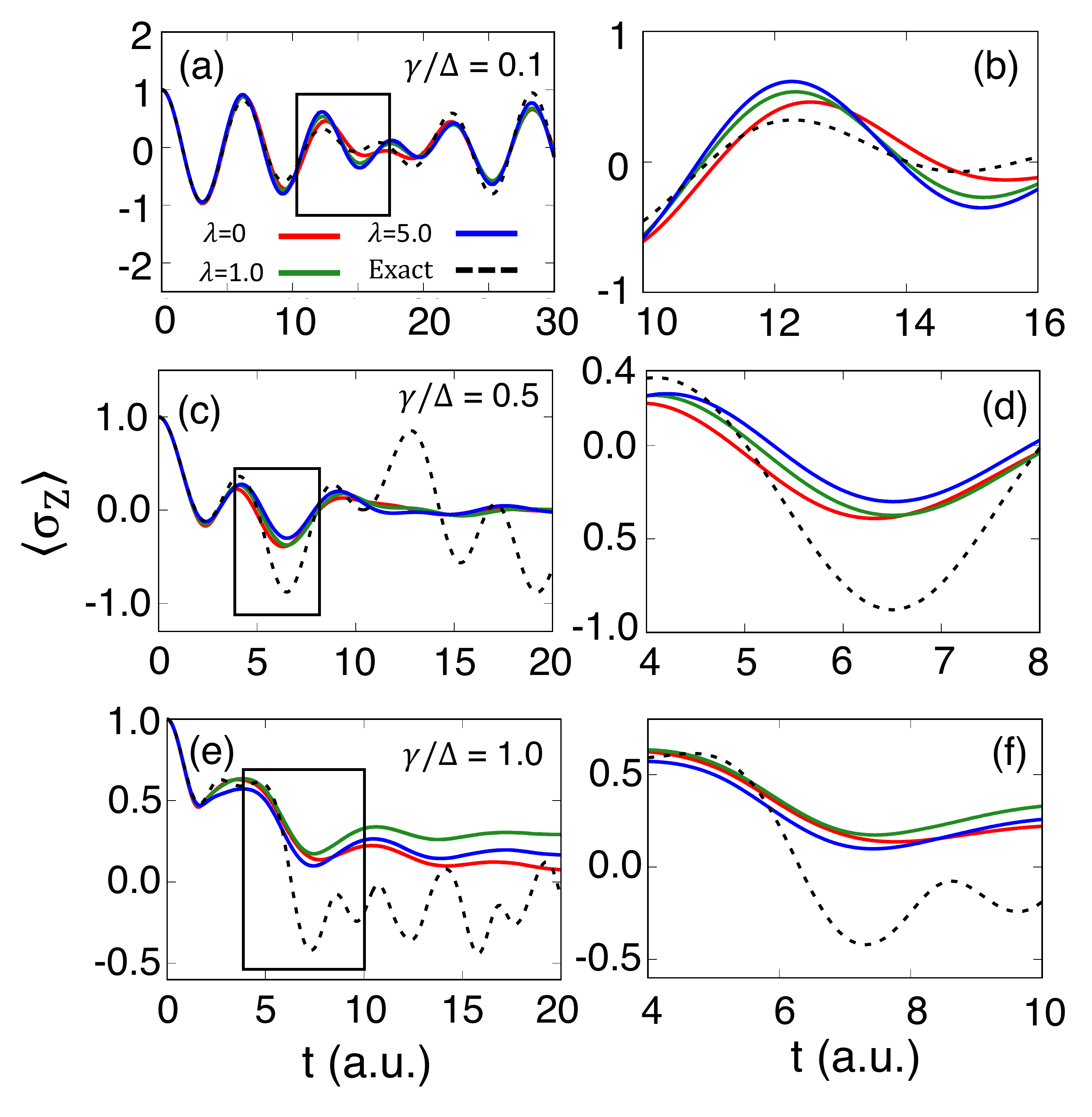}
  \end{minipage}%
   \caption{Population dynamics of Model II, with the left column shows $\langle\sigma_Z\rangle$, and the right column depicts the magnified plots that correspond to the square regions in the left column. Results are obtained from T-NRPMD with different friction parameters, $\lambda=0$ (red), $\lambda=1$ (green), $\lambda=5$ (blue),  as well as the numerical exact results (black dashed line).}
\label{fig:trpmd}
\end{figure}

Fig.~\ref{fig:trpmd} presents the results of Model II obtained from T-NRPMD. Here, we investigate the effects of nuclear thermostatting on the excited state non-adiabatic dynamics and explore the impact of various friction constant $\lambda$ on the population dynamics. The results are obtained by using $\lambda=0$ (red), {\it i.e.}, the NRPMD approach (same as results shown in Fig.~\ref{fig:1dspinboson}), $\lambda=1.0$ (green), and $\lambda=5.0$ (blue), together with the numerically exact results (black dashed lines). Despite that NRPMD provides more accurate population dynamics compared to classical Wigner based methods as discussed in Fig.~\ref{fig:1dspinboson}, the population tends to oscillate with a slightly shifted frequency compared to the electronic Rabi frequency. This can be clearly seen by comparing the results of NRPMD (red) and the exact ones (black) in panels (b),(d) and (f) which depict the magnified plots corresponding to the square regions in panels (a), (c), and (e). The higher frequency normal modes of the nuclear ring polymer might be the source of these spurious oscillations,\cite{hele2016_trpmd} which has shown to contaminate the nuclear dynamics.\cite{hele2016_trpmd}

Using $\lambda=1.0$ (green), T-NRPMD recovers the correct oscillation frequencies of the electronic population. This value of $\lambda$ is chosen based on a friction parameter that is derived from achieving the correct nuclear oscillation frequency in a harmonic potential.\cite{hele2016_trpmd} By recovering the correct nuclear oscillations of $R$, the coupled electronic dynamics is also improved. We also observed that by applying a small friction parameter $0.5\leqslant\lambda\leqslant1$, T-NRPMD already improves the dynamics and recover the correct oscillation period. The population dynamics continues to oscillate with the correct frequency when further increase the friction parameter to the overdamped regime with $\lambda=5.0$ (solid blue line). The correct electronic oscillation is likely stemmed from the correct nuclear oscillations when applying thermostat which has been demonstrated in the previous work.\cite{hele2016_trpmd} 

These investigations demonstrate that T-NRPMD is a valuable tool for providing accurate excited state dynamics and alleviate spurious frequency problem associated with the ring polymer quantization. Future studies include rigorous derivation of T-NRPMD approach through the Matsubara dynamics framework\cite{hele2015,hele2015jcp2} with the mapping Liouvillian\cite{hele2016} to treat electronic states explicitly. 

\bibliographystyle{aipnum4-1}
\bibliography{wm-rpmd_v15.bbl}
\end{document}